\documentclass[letter,usenatbib]{mn2e}
\usepackage{epsfig}
\usepackage{times}
\def\gtrsim{~\rlap{$>$}{\lower 1.0ex\hbox{$\sim$}}}
\def\ltsim{~\rlap{$<$}{\lower 1.0ex\hbox{$\sim$}}}

\title[Relations among 8, 24, and 160~$\mu$m dust emission]
    {The relations among 8, 24, and 160~$\mu$m dust emission within nearby 
    spiral galaxies}
\author[G. J. Bendo et al.]
    {G. J. Bendo$^1$, B. T. Draine$^2$, C. W. Engelbracht$^3$, G. Helou$^4$, 
     M. D. Thornley$^5$, \newauthor 
     C. Bot$^6$, B. A. Buckalew$^7$, D. Calzetti$^8$, 
     D. A. Dale$^9$, D. J. Hollenbach$^{10}$,  \newauthor 
     A. Li$^{11}$, J. Moustakas$^{12}$ \\
    $^1$Astrophysics Group, Imperial College, Blackett Laboratory,
        Prince Consort Road, London SW7 2AZ, United Kingdom\\
    $^2$Princeton University Observatory, Peyton Hall, Princeton, 
        NJ 08544-1001, USA\\
    $^3$Steward Observatory, University of Arizona, 933 North 
        Cherry Avenue, Tucson, AZ 85721, USA\\
    $^4$California Institute of Technology, MC 314-6, Pasadena, CA 
        91101, USA\\
    $^5$Department of Physics \& Astronomy, Bucknell University, 
        Lewisburg, PA 17837, USA\\
    $^6${\it Spitzer} Science Center, California Institute of Technology, 
        MS 220-6, Pasadena, CA 91101, USA\\
    $^7$Department of Physics, Embry-Riddle Aeronautical University, 3700 
        Willow Creek Road, Prescott, AZ 86301, USA\\
    $^8$Department of Astronomy, University of Massachusetts, 
        LGRT-B 254, 710 North Pleasant Street, Amherst, MA 01002, USA\\
    $^9$Department of Physics and Astronomy, University of Wyoming, 
        Laramie, WY 82071, USA\\
    $^{10}$NASA Ames Research Center, MS 245-3, Moffett Field, CA 
        94035-1000, USA\\
    $^{11}$Department of Physics and Astronomy, University of 
        Missouri, Columbia, MO 65211, USA\\
    $^{12}$Department of Physics, New York University, 4 Washington Place, New
        York, NY 10003, USA\\
}
\date{}
\pagerange{\pageref{firstpage}--\pageref{lastpage}}
\pubyear{}

\begin{document}
\label{firstpage}
\maketitle

\begin{abstract}
We investigate the relations among the stellar continuum-subtracted
8~$\mu$m polycyclic aromatic hydrocarbon (PAH 8~$\mu$m) emission,
24~$\mu$m hot dust emission, and 160~$\mu$m cold dust emission in
fifteen nearby face-on spiral galaxies in the {\it Spitzer} Infrared
Nearby Galaxies Survey sample.  The relation between PAH 8 and
24~$\mu$m emission measured in $\sim2$~kpc regions is found to exhibit
a significant amount of scatter, and strong spatial variations are
observed in the (PAH 8~$\mu$m)/24~$\mu$m surface brightness ratio.  In
particular, the (PAH 8~$\mu$m)/24~$\mu$m surface brightness ratio is
observed to be high in the diffuse interstellar medium and low in
bright star-forming regions and other locations with high 24~$\mu$m
surface brightness.  PAH 8~$\mu$m emission is found to be
well-correlated with 160~$\mu$m emission on spatial scales of
$\sim2$~kpc, and the (PAH 8~$\mu$m)/160~$\mu$m surface brightness
ratio is generally observed to increase as the 160~$\mu$m surface
brightness increases.  These results suggest that the PAHs are
associated with the diffuse, cold dust that produces most of the
160~$\mu$m emission in these galaxies, and the variations in the (PAH
8~$\mu$m)/160~$\mu$m ratio may generally be indicative of either the
intensity or the spectrum of the interstellar radiation field that is
heating both the PAHs and the diffuse interstellar dust.
\end{abstract}


\section{Introduction \label{s_intro}}

Observations of nearby spiral galaxies with the Infrared Space
Observatory (ISO) demonstrated that polycyclic aromatic hydrocarbon
(PAH) spectral feature emission at $\sim8$~$\mu$m is closely
associated with hot dust emission at 15~$\mu$m, and the 8~$\mu$m
emission was also shown to be correlated with other star formation
tracers \citep{retal01, frsc04}.  However, observations with the {\it
Spitzer} Space Telescope \citep{wetal04} have demonstrated that the
correlation of PAH emission at 8~$\mu$m to hot dust emission at
24~$\mu$m was not necessarily a one-to-one correlation within
individual galaxies.  \citet{hraetal04}, \citet{betal06}, and
\citet{getal08} demonstrated that, on scales of hundreds of parsecs,
continuum-subtracted PAH emission at 8~$\mu$m (henceforth referred to
as PAH 8~$\mu$m emission) does not accurately trace 24~$\mu$m
emission.  In particular, PAH 8~$\mu$m emission seems to appear in
shell-like features around star-forming regions, while 24~$\mu$m
emission peaks within star-forming regions.  Using images of M51,
\citet{cetal05} demonstrated that PAH 8~$\mu$m emission compared to
other tracers of star formation, including 24~$\mu$m emission, is
disproportionately low within star-forming regions and
disproportionately high in diffuse regions.  The results from
\citet{cetal07} show that the PAH 8~$\mu$m/24~$\mu$m emission ratio
measured in H{\small II} regions is partly dependent on the aperture
used for measuring the flux densities, which also suggest that a
significant fraction of PAH 8~$\mu$m emission originates from outside
the H{\small II} regions.  \citet{tetal07} also demonstrated that the
PAH 8~$\mu$m/24~$\mu$m ratio of H{\small II} regions in NGC~7331 may
depend on infrared, H$\alpha$, or ultraviolet surface brightness after
corrections for dust extinction have been applied, which implied that
PAH 8~$\mu$m emission is suppressed within strong star-forming
regions.  \citet{eetal05}, \citet{detal05}, \citet{ddbetal07}, and
\citet{eetal08} have demonstrated that the global 8~$\mu$m/24~$\mu$m
emission ratio may decrease in regions with lower metallicity, and
\citet{cetal07} also demonstrated that this ratio for H{\small II}
regions in other galaxies depends on metallicity.  Although these
studies based on {\it Spitzer} data have shown explicitly or
implicitly that the relation between 8~$\mu$m PAH emission and
24~$\mu$m emission on $\ltsim 1$~kpc scales is not a one-to-one
relation, no study has yet explored variation in the (PAH
8~$\mu$m)/24~$\mu$m ratio between diffuse and star-forming regions
within a broad range of spiral galaxies.

While most research has focused on comparisons between PAH emission
and hot dust emission, some additional studies have compared PAH
emission to cold dust emission at wavelengths longer than 100~$\mu$m.
Using ISO data, \citet{mll99} and \citet{hkb02} demonstrated that PAH
emission at 8~$\mu$m is correlated with large grain emission, and
\citet{hkb02} even argued that the correlation between 8~$\mu$m
emission and cold dust emission at 850~$\mu$m was much stronger than
the correlation between 8 and 15~$\mu$m emission.  Using {\it Spitzer}
data, \citet{betal06} demonstrated that PAH 8~$\mu$m emission was
correlated with 160~$\mu$m cold dust emission in NGC~4631, but the
ratio of (PAH 8~$\mu$m)/160~$\mu$m emission varied.  Although the (PAH
8~$\mu$m)/160~$\mu$m ratio appeared related to the infrared surface
brightness in NGC~4631, \citet{betal06} used bright infrared sources
in the outer disc to argue that the variations in the (PAH
8~$\mu$m)/160~$\mu$m ratio are instead dependent on radius.  Given
that the ratio of PAH emission to dust emission at longer wavelengths
or the ratio of PAHs to total dust mass may vary with metallicity
\citep{eetal05, detal05, setal07, ddbetal07, eetal08} and that
metallicity is generally expected to vary with radius within spiral
galaxies \citep[e.g.][J. Moustakas et al. 2008, in
preparation]{s71, ws83, ve92, zkh94, vetal98, petal04}, the results
from NGC~4631 suggested that the variations in (PAH
8~$\mu$m)/160~$\mu$m ratio within that galaxy could be related to
metallicity.  Unfortunately, NGC~4631 is viewed edge-on, so local
variations in the observed (PAH 8~$\mu$m)/160~$\mu$m ratio that are
not dependent on radius may be suppressed by the line-of-sight
integrations through the disc.  Variations in the (PAH
8~$\mu$m)/160~$\mu$m ratio and its dependence on both infrared surface
brightness and radius need to be studied further using spiral galaxies
with orientations closer to face-on.

In this paper, we examine the relation of PAH 8~$\mu$m emission to
both 24~$\mu$m hot dust emission and 160~$\mu$m cold dust emission in
a set of fifteen face-on spiral galaxies observed by {\it Spitzer} as
part of the {\it Spitzer} Infrared Nearby Galaxies Survey (SINGS)
legacy project \citep{kabetal03}.  The basic goal is to understand how
PAH 8~$\mu$m emission is related to dust emission at other
wavelengths.  Section~\ref{s_data} provides basic information on the
wave bands used in the analysis, the observations and data reductions,
the sample, and the preparation of the data for the analysis.  The
comparison between PAH 8 and 24~$\mu$m emission is presented in
Section~\ref{s_comp_pah24}, and the comparison between PAH 8 and
160~$\mu$m emission is presented in Section~\ref{s_comp_pah160}.  The
analysis is then followed by a discussion in Section~\ref{s_discuss}
and conclusions in Section~\ref{s_conclusions}.

\section{Data}
\label{s_data}

\subsection{Wave band information}
\label{s_data_band}

Here we present background information on the 8~$\mu$m, 24~$\mu$m, and
160~$\mu$m bands used for the analysis in this paper.  This
information also includes some caveats that should be considered when
interpreting the results from these data.

Channel 4 of the Infrared Array Camera \citep[IRAC;][]{fetal04} covers
a region centered on 8.0~$\mu$m that includes the 7.7~$\mu$m PAH
feature as well as some stellar continuum and some hot dust emission
\citep[e.g.][]{setal07}.  For the analysis here, the stellar continuum
is subtracted using 3.6~$\mu$m data from channel 1 of IRAC.  Although
the PAH emission features are very prominent in this 8~$\mu$m band,
thermal emission from larger grains may contribute significantly to
emission in this band in regions with very intense radiation fields
\citep{dl07}.

The 24~$\mu$m detector of the Multiband Imaging Photometer for {\it
Spitzer} \citep[MIPS;][]{ryeetal04} mostly detects hot
($\gtrsim100$~K) dust emission within nearby galaxies.  In
environments with low radiation fields, this dust emission may
originate from mostly transiently-heated small grains, but in regions
with high radiations fields such as star-forming regions, the emission
consists of mostly thermal emission from grains at equilibrium
temperatures of $\gtrsim100$~K \citep[e.g.][]{ld01, dl07}.  The
relative contribution of dust at $\gtrsim100$~K is expected to be
strongly dependent on the strength of the illuminating radiation
field.  Consequently, the 24~$\mu$m band is predicted to increase more
rapidly than other infrared bands as the illuminating radiation field
increases, as has been shown with many physical and semi-empirical
models of dust emission \citep[e.g.][]{dhcsk01, ld01, dl07}.  Multiple
studies with {\it Spitzer} data have shown that the 24~$\mu$m emission
from point-like sources corresponds to H{\small II} regions visible in
optical and ultraviolet wave bands, so the 24~$\mu$m band may be used
to measure star formation activity \citep{pkbetal07, cetal07}.
Extended 24~$\mu$m emission, however, may originate from outside of
star-forming regions in the diffuse interstellar medium.

The 160~$\mu$m MIPS detector mainly traces cold ($\sim20$~K) dust
emission that may be associated with cirrus dust in the diffuse
interstellar medium \citep[e.g.][]{ddbetal07}.  Because the 160~$\mu$m
band samples dust emission near the peak of the spectral energy
distribution in the galaxies in this sample \citep{dggetal07}, the
160~$\mu$m band should be tightly correlated with the total infrared (TIR)
luminosity.  This is explained in the following proof.  For thermal
emission modified by an emissivity function that scales as
$\nu^\beta$, the total energy emitted in a single wave band ($\nu
L_\nu$) should scale as
\begin{equation}
\nu L_\nu \propto \frac{\nu^{(4+\beta)}}{e^x-1}
\end{equation}
where 
\begin{equation}
x=\frac{h\nu}{kT}.
\end{equation}
The total integrated thermal emission $L_{total}$ will scale according
to
\begin{equation}
L_{total} \propto T^{(4+\beta)}.
\end{equation}
The ratio of $\nu L_\nu$ to $L_{total}$ will be approximately constant
(e.g. independent of grain temperature) when it is close to a maximum,
which occurs when
\begin{equation}
\frac{xe^x}{e^x-1}=4+\beta.
\end{equation}
Since $e^x >> 1$, then
\begin{equation}
T \approx \frac{1}{4+\beta}\frac{h\nu}{k}.
\end{equation}
Assuming that $\beta$ equals 2 \citep{ld01}, this condition is met at
160~$\mu$m when the dust temperature is $\sim15$~K.  Because the peak
of the spectral energy distribution is actually at a shorter
wavelength than 160~$\mu$m, the ratio of 160~$\mu$m emission to total
infrared emission should decrease slightly as the illuminating
radiation field increases.

\subsection{Observations and data reduction}

We use the 3.6, 8.0, 24, 70, and 160~$\mu$m data taken with {\it
Spitzer} as part of SINGS.  The 3.6 and 8.0~$\mu$m observations were
performed with IRAC.  The observations for each object consisted of a
series of 5~arcmin $\times$ 5~arcmin individual frames taken in either
a mosaic or a single field dither pattern.  The 24, 70, and 160~$\mu$m
observations performed with MIPS are composed of two scan maps for
each target.  Each object was observed twice in each wave band to
identify and remove transient phenomena, particularly asteroids.  The
full-width half-maxima (FWHM) of the point spread functions (PSFs), as
stated in the Spitzer Observer's Manual
\citep{sscmanual06}\footnote[6]{http://ssc.spitzer.caltech.edu/documents/som/},
are 1.7, 2.0, 6, and 38~arcsec at 3.6, 8.0, 24, and 160~$\mu$m,
respectively.  Details on the observations can be found in the
documentation for the SINGS fourth data delivery
\citep{sings06}\footnote[7]{Available at
http://ssc.spitzer.caltech.edu/legacy/singshistory.html}.

The IRAC data were processed using the SINGS IRAC pipeline, which
combines multiple frames of data using a drizzle technique.  A
description of the technique is presented in \citet{retal06}.  The
final images may contain residual background emission from the
telescope or sky that is subtracted during the analysis.  The MIPS
data were processed using the MIPS Data Analysis Tools version 3.06
\citep{getal05}.  Additional software was used to remove zodiacal light
emission and improve the flatfielding in the 24~$\mu$m data and to
remove short-term variations in the background signal (commonly
referred to as drift) in the 70 and 160~$\mu$m data.  Any additional
background offset left in the final images was measured in regions
outside the optical discs and subtracted.  Additional details are
presented in \citet{betal06}.  Full details on the data processing are
also available in the SINGS documentation for the fourth data delivery
\citep{sings06}.

\subsection{Sample selection}
\label{s_sample}

To perform this analysis, we need to resolve substructures in the
160~$\mu$m images, which have PSFs with FWHM of 38~arcsec.  We
therefore limit the sample to spiral galaxies in SINGS where the major
axes of the D$_{25}$ isophote specified by \citet{ddcbpf91} are larger
than 5~arcmin.  Since we want to be able to distinguish between radial
colour variations and colour variations related to the presence of
substructures and since such substructures are difficult to study in
edge-on galaxies, we only use galaxies that are inclined less than
$\sim60\deg$.  The inclinations are calculated using
\begin{equation}
i=\cos^{-1}\left( \sqrt{\frac{q^2-q_o^2}{1-q_o^2}} \right).
\label{e_inclination}
\end{equation} 
The value $q$ is the observed (projected) minor-to-major axis ratio.
The value $q_o$ is the intrinsic optical axial ratio (the ratio of the
unprojected optical axis perpendicular the plane of the galaxy to the
diameter of the disc), which is equivalent to 0.20 for most disc
galaxies \citep{t88}.  Because the optical disc of NGC~5194 may be
distorted by its interaction with NGC~5195, its inclination is not
calculated using this equation.  Instead, the inclination as well as
the position angle given by \citet{gab02} are used for the analysis.

Six of the SINGS galaxies that meet the above criteria are unsuitable
for the analysis.  NGC~1512 and NGC~4826 are not used because only the
central regions were detected at the $5\sigma$ levels in all of the
convolved maps (described in the next section).  The 8.0~$\mu$m images
of NGC~1097, NGC~1566, and NGC~4736 are heavily affected by muxbleed
artefacts (artificially bright columns of pixels associated with
high-surface brightness sources) that cross over significant fractions
of the optical discs, so those data are not usable for this analysis.
Two very bright foreground stars in the 3.6 and 8.0~$\mu$m images of
NGC~3621 cause problems in the analysis, so NGC~3621 needs to be
excluded from the sample as well.  The other 15 galaxies that meet the
above criteria, which are roughly uniformly distributed between Hubble
types Sab and Sd, are listed in Table~\ref{t_sample} along with
information on the galaxies' morphologies, optical axes, distances,
nuclear spectral types, and nebular oxygen abundances (12+log(O/H),
which is treated as representative of the global metallicities of the
galaxies).

\begin{table*}
\centering
\begin{minipage}{168mm}
\caption{Basic Properties of the Sample Galaxies \label{t_sample}}
\begin{tabular}{@{}lcccccccc@{}}
\hline
Name &        
    Hubble &            
    Size of Optical &
    Inclination$^b$ &
    Position &    
    Distance &
    Distance &
    Nuclear &
    12+log(O/H)$^g$ \\
&             
    Type$^a$ &
    Disc (arcmin)$^a$ 
    &
    Angle$^c$ &
    (Mpc)$^d$ &
    &
    Reference$^e$ &
    Type$^f$ &
    \\
\hline
NGC 628 &     SA(s)c &            $10.5 \times 9.5$ &    
    $26^\circ$ &      $25^\circ$ &    
    $7.3 \pm 1.4$ &   1 &
    SF &              $8.33 \pm 0.02$ \\
NGC 925 &     SAB(s)d &           $10.5 \times 5.9$ &
    $58^\circ$ &      $102^\circ$ &    
    $9.12 \pm 0.17$ & 2 &
    SF &              $8.24 \pm 0.01$ \\
NGC 2403 &    SAB(s)cd &          $21.9 \times 12.3$ &
    $58^\circ$ &      $127^\circ$ &            
    $3.13 \pm 0.14$ & 2 &
    SF &              $8.31 \pm 0.01$ \\
NGC 3031 &    SA(s)ab &           $26.9 \times 14.1$ &
    $60^\circ$ &      $157^\circ$ &            
    $3.55 \pm 0.15$ & 2 &
    AGN &             $8.40 \pm 0.01$ \\
NGC 3184 &    SAB(rs)cd &         $7.4 \times 6.9$ &     
    $22^\circ$ &      $135^\circ$ &            
    $11.1 \pm 1.9$ &  3 &
    SF &              $8.48 \pm 0.02$ \\
NGC 3351 &    SB(r)b &            $7.4  \times 5.0$ &
    $49^\circ$ &      $13^\circ$ &             
    $9.3 \pm 0.4$ &   2 &
    SF &              $8.61 \pm 0.01$ \\
NGC 3938 &    SA(s)c &            $5.4 \times 4.9$ &     
    $25^\circ$ &      $42^\circ\ast$ &         
    $13.4 \pm 2.3$ &  4 &
    SF &              $8.44 \pm 0.18$ \\
NGC 4254 &    SA(s)c &            $5.4 \times 4.7$ &     
    $30^\circ$ &      $68^\circ\ast$ &         
    $16.5 \pm 0.6$ &  5 &
    SF &              $8.46 \pm 0.02$ \\
NGC 4321 &    SAB(s)bc &          $7.4 \times 6.3$ &     
    $32^\circ$ &      $30^\circ $&             
    $14.3 \pm 0.5$ &  2 &
    AGN &             $8.50 \pm 0.04$ \\
NGC 4579 &    SAB(rs)b &          $5.9 \times 4.7$ &
    $38^\circ$ &      $95^\circ$ &             
    $16.5 \pm 0.6$ &  5 &
    AGN &             $8.55 \pm 0.18$ \\
NGC 4725 &    SAB(r)ab pec &      $10.7 \times 7.6$ &
    $46^\circ$ &      $35^\circ$ &             
    $11.9 \pm 0.3$ &  2 &
    AGN &             $8.40 \pm 0.05$ \\
NGC 5055 &    SA(rs)bc &          $12.6 \times 7.2$ &
    $57^\circ$ &      $105^\circ$ &            
    $7.8 \pm 2.3$ &   4  &
    AGN &             $8.42 \pm 0.04$ \\
NGC 5194 &    SA(s)bc pec &       $11.2 \times 6.9$ &
    $20^\circ$ &      $170^\circ$ &            
    $8.4 \pm 0.6$ &   6 &
    AGN &             $8.54 \pm 0.01$ \\
NGC 6946 &    SAB(rs)cd &         $11.5 \times 9.8$ &
    $32^\circ$ &      $80^\circ\ast$ &         
    $6.8 \pm 1.7$ &   7 &
    SF &              $8.40 \pm 0.04$ \\
NGC 7793 &    SA(s)d &            $9.3 \times 6.3$ &
    $49^\circ$ &      $98^\circ$ &             
    $3.9 \pm 0.4$ &   8 &
    SF &              $8.27 \pm 0.03$ \\
\hline
\end{tabular}
$^a$ These data are taken from \citet{ddcbpf91}.  The optical disc
     is the size of the D$_{25}$ isophote. \\
$^b$ The inclinations are calculated using the dimensions of the galaxies and 
     Equation~\ref{e_inclination}. An exception is made for NGC~5194, where
     the inclination from \citet{gab02} was used.\\
$^c$ The position angles are measured in degrees from north through
     east.  These data are taken from \citet{ddcbpf91} when given.
     The position angle for NGC~5194 was taken from \citet{gab02}.
     Position angles marked with an asterisk are measured in the
     3.6~$\mu$m data using the IDL fit\_ellipse program by D. Fanning
     (accessible from
     http://www.dfanning.com/documents/programs.html).  These position
     angles have uncertainties of $\sim5^\circ$. \\
$^d$ These distances are part of a compilation of distance for the 
     complete SINGS sample will be presented in J. Moustakas et al. (2008, 
     in preparation).\\
$^e$ Distance references - (1) \cite{skt96}; (2) \cite{fetal01};
     (3) \cite{letal02}; (4) \cite{m05}; (5) \cite{metal07}; (6) \cite{fcj97};
     (7) \cite{ksh00}; (8) \cite{kgsetal03}.\\
$^f$ The nucleus types are based on the analysis of the optical
     spectrophotometric properties by J. Moustakas et al. (2008, in
     preparation).  The spectral type indicates whether the center of
     each object is dominated by star formation (SF) or an active
     galactic nucleus (AGN).  See J. Moustakas et al. (2008, in
     preparation) for more details.\\
$^g$ These are the characteristic nebular oxygen abundances presented
     by J. Moustakas et al. (2008, in preparation) using the
     \citet{pt05} strong-line calibration.  They are representative of
     the mean, luminosity-weighted oxygen abundance of each galaxy.
     Note that the uncertainties only include statistical measurement
     errors, and do not include any systematic uncertainties in the
     adopted abundance scale (see J. Moustakas et al. 2008, in
     preparation for more details).  Although J. Moustakas et al. also
     present abundances based on the \citet{kk04} strong-line
     calibration, we adopt the abundances based on the \citet{pt05}
     calibration because they are comparable to the abundances used by
     \citet{eetal08}.
\end{minipage}
\end{table*}

\subsection{Data preparation}
\label{s_dataprep}

Many {\it Spitzer} studies of infrared colour variations within
individual galaxies have relied on flux densities measured within
discrete subregions that are chosen either by eye or by source
identification software.  However, this selection process may be
biased.  Some subregions within the galaxy might be excluded from the
analysis, especially if the regions are selected by eye, and region
selection may also be biased towards regions that appear bright in a
specific wave band.  To avoid this bias, we divide each galaxy into
45~arcsec square regions and extract surface brightnesses for all
usable regions within the optical discs of the galaxies.  This
approach samples the whole of the galactic disc and avoids any
subjective biases.  We choose to use 45~arcsec regions because it is
an integer multiple of the pixels used in the original images, and it
is larger than the 38~arcsec FWHM of the 160~$\mu$m images.  Note that
45~arcsec corresponds to physical scales of $\sim0.7$-3.6~kpc for the
galaxies in this sample.

First, the data are convolved with kernels that match the PSFs of the
images in the 3.6, 8, 24, and 70~$\mu$m bands to the PSF of the
160~$\mu$m data, which has a FWHM of 38~arcsec.  The convolution
kernels were created by \citep{getal08} using the ratio of the Fourier
transforms of the input and output PSFs, with high-frequency noise
suppressed in the input PSFs\footnote[8]{\raggedright The kernels are
available at
http://dirty.as.arizona.edu/$\sim$kgordon/mips/conv\_psfs/conv\_psfs.html.}.
By doing this, it is possible to directly compare surface brightnesses
measured in the same apertures in different wave bands without
applying any aperture corrections.  Next, the coordinate systems
across all the wave bands are matched to each other using point-like
sources (stars, background galaxies, or infrared-bright regions within
the galaxies) or the centres of the galaxies as guides.  The
background in the IRAC data was then subtracted from the images.
Following this, the data are rebinned into 45~arcsec square regions
such that the coordinates of the rebinned pixels match across all wave
bands.  Because the centres of the galaxies are mapped into the
corners of four pixels in the MIPS images, the centres of the galaxies
fall at the corners of four 45~arcsec bins.  We then extract surface
brightnesses from all 45~arcsec regions within the optical discs of
the galaxies.  We exclude from the analysis regions not detected at
the $3\sigma$ level in one or more wave bands, regions contaminated by
emission from bright foreground stars (identified as unresolved
sources in the unconvolved IRAC data with 3.6~$\mu$m/8~$\mu$m surface
brightness ratios $\gtrsim 5$), and regions affected by muxbleed at
8~$\mu$m.  Additionally, we exclude the centre of NGC~5195 from the
analysis of NGC~5194 because the dwarf galaxy causes confusion when
interpreting the data on the spiral galaxy.  To correct for the
diffusion of light through the IRAC detector substrate, the 3.6 and
8.0~$\mu$m data are multiplied by the ``infinite'' aperture
corrections described by \citet{retal05} (which should not be confused
with the types of aperture corrections used for measuring the flux
densities of unresolved sources).  The correction factors are 0.944
and 0.737 at 3.6 and 8.0~$\mu$m, respectively.  Finally, we subtract
the stellar continuum from the 8 and 24~$\mu$m surface brightnesses
(in MJy sr$^{-1}$) using
\begin{equation}
I_\nu(PAH~8\mu m)=I_\nu(8\mu m)-0.232I_\nu(3.6\mu m)
\label{e_8starsub}
\end{equation}
\begin{equation}
I_\nu(24\mu m)=I_\nu(24\mu m)-0.032I_\nu(3.6\mu m),
\label{e_24starsub}
\end{equation}
which were derived by \citet{hraetal04}.  For the galaxies in this
analysis, this is a correction of $\sim1$~\%-4\% to the 24~$\mu$m
surface brightnesses but a correction of $\sim5$\%-25\% to the
8~$\mu$m surface brightnesses.  The stellar continuum-subtracted
8~$\mu$m data are referred to as the PAH 8~$\mu$m emission throughout
this paper.  While these stellar continuum-subtracted spectra may
contain some thermal dust emission, most of the emission should
originate from the PAH emission features, as has been demonstrated by
\citet{setal07}.  We also calculated TIR surface brightnesses using 
\begin{equation}
\begin{array}{l}
I(TIR)=0.95\nu I_\nu(PAH 8~\mu m)+1.15\nu I_\nu(24~\mu m)\\
+\nu I_\nu(70~\mu m)+\nu I_\nu(160~\mu m)
\end{array}
\end{equation}
based on equation 22 from \citet{dl07}.

In the analysis of these 45~arcsec regions, we assume that the
background noise, which is measured in off-target regions in the
images, is the only source of uncertainties that will affect this
analysis.  Uncertainties in the calibration will only scale the data,
which does not affect either the slopes of lines fit to the data in
log space or the scatter around the best fit lines.  Uncertainties in
the mean background value subtracted from the data should be
negligible.  The uncertainties for the ratios of surface brightnesses
in two wave bands also includes a term to account for uncertainties in
matching the coordinate systems of the two bands.  This term is
estimated from the standard deviations of the central four pixels in
galaxies with symmetric, bright, point-like nuclei.  In galaxies with
such nuclei, the central four pixels should each sample approximately
one-quarter of the peak of the central PSF.  They should have the same
surface brightness ratios if the image coordinate systems of different
images are matched properly, but in practice, the ratios vary by small
amounts.  These terms, which are given in
Table~\ref{t_ratiouncertainty} for the ratios used in this paper, add
uncertainties of approximately 5\%-10\%.

\begin{table}
\begin{center}
\renewcommand{\thefootnote}{\alph{footnote}}
\caption{Uncertainties in the Surface Brightness Ratios Related to
Matching Image Coordinate Systems \label{t_ratiouncertainty}}
\begin{tabular}{@{}cc@{}}
\hline
Surface &                                   Uncertainty \\
Brightness &                                \\
Ratio &                                     \\
\hline
$I_\nu$(PAH 8~$\mu$m)/$I_\nu$(24~$\mu$m) &  0.005 \\
$I_\nu$(PAH 8~$\mu$m)/$I_\nu$(160~$\mu$m) & 0.0015 \\
$\nu I_\nu$(PAH 8~$\mu$m)/$I$(TIR) &        0.007 \\
$I\nu$(24~$\mu$m)/$I_\nu$(160~$\mu$m) &     0.005 \\
$\nu I_\nu$(160~$\mu$m)/$I$(TIR) &          0.02 \\
\hline
\end{tabular}
\end{center}
\end{table}

We also use maps of the (PAH 8~$\mu$m)/24~$\mu$m and (PAH
8~$\mu$m)/160~$\mu$m surface brightness ratios as additional aids in
the interpretation of the data.  To make these ratio maps, we again
use the convolution kernels of K. D. Gordon to match the PSFs of the
3.6 and 8.0~$\mu$m images to the PSF in the longest wave band.  Hence,
the 3.6, 8.0, and 24~$\mu$m images used to produce the (PAH
8~$\mu$m)/24~$\mu$m maps have PSFs with FWHM of 6~arcsec, and the 3.6,
8.0, and 160~$\mu$m images used to produce the (PAH
8~$\mu$m)/160~$\mu$m maps have FWHM of 38~arcsec.  The coordinate
matching, IRAC background subtraction, IRAC aperture correction, and
stellar continuum subtraction steps applied to the surface
brightnesses measured in the 45~arcsec regions are also applied to the
images used to make these ratio maps.  Only pixels where the 24~$\mu$m
data were detected at the $5\sigma$ level and where the PAH 8 and
160~$\mu$m data were detected at the $10\sigma$ level are shown in the
maps.  These thresholds were selected to filter out background noise
and artefacts, which is why the threshold differs among the three wave
bands.  Moreover, pixels above these thresholds roughly correspond to
the parts of the disc that were binned into the 45~arcsec regions
described above.  Additionally, foreground stars and any bright
artefacts that remained in the data were masked out in the final ratio
maps.  Images of the PAH 8~$\mu$m emission, the (PAH
8~$\mu$m)/24~$\mu$m surface brightness ratio, and the (PAH
8~$\mu$m)/160~$\mu$m surface brightness ratio are shown in
Figure~\ref{f_map}.

\begin{figure*}
\begin{center}
\epsfig{file=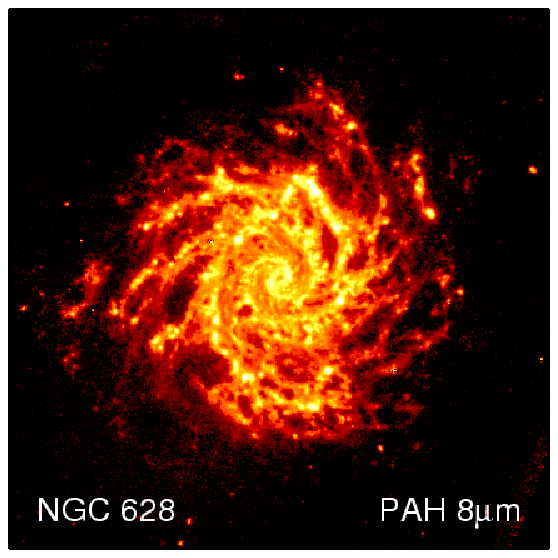, height=50mm}
\epsfig{file=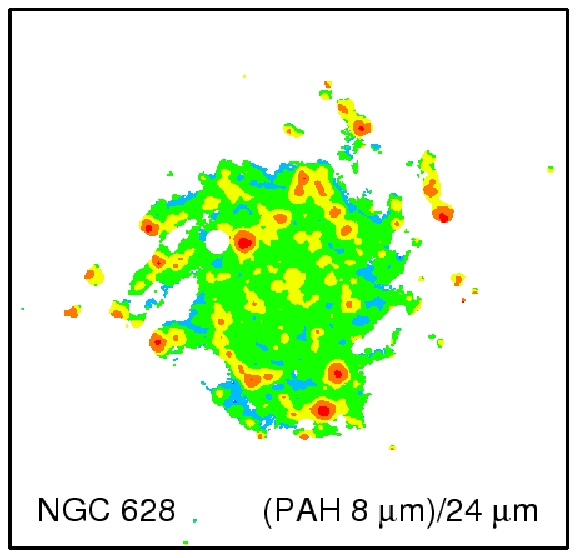, height=50mm}
\epsfig{file=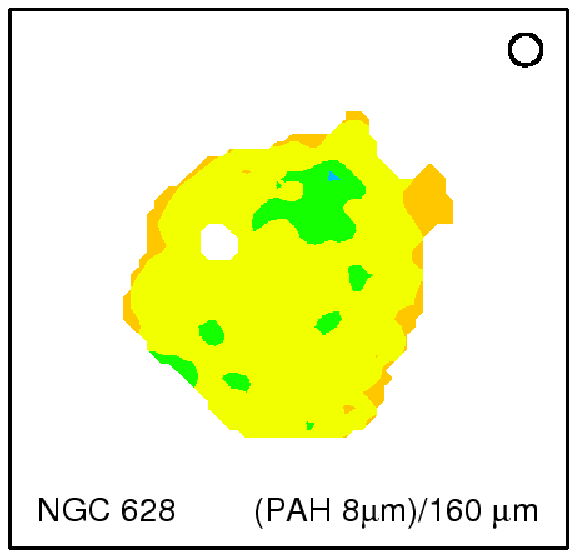, height=50mm}
\epsfig{file=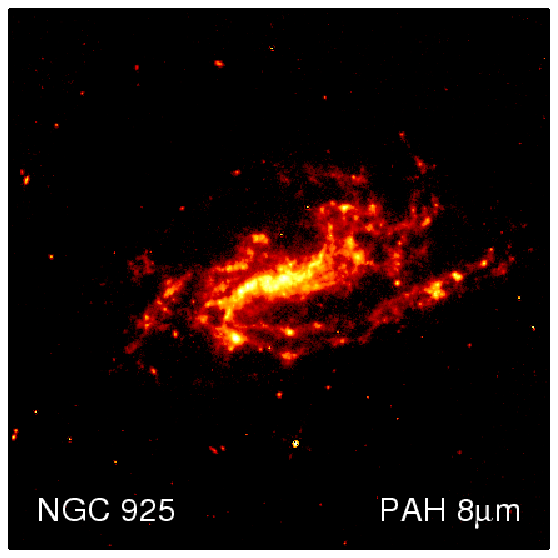, height=50mm}
\epsfig{file=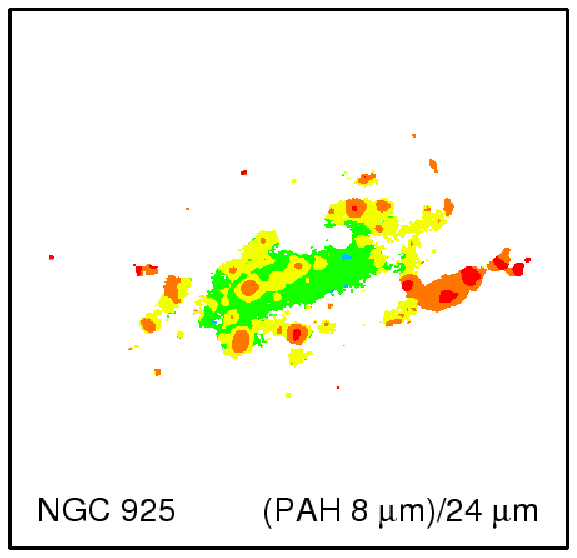, height=50mm}
\epsfig{file=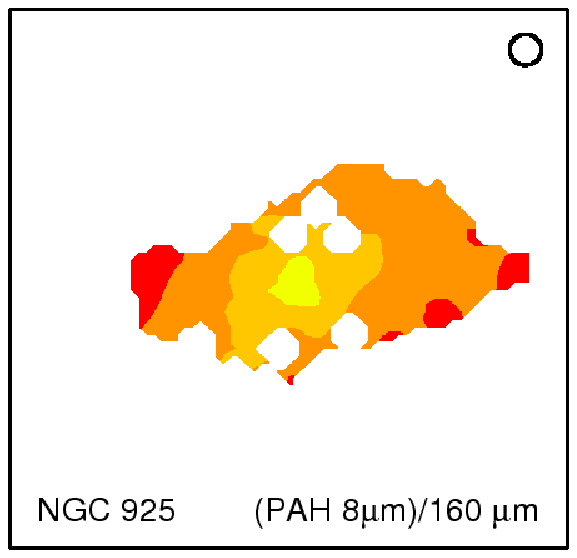, height=50mm}
\epsfig{file=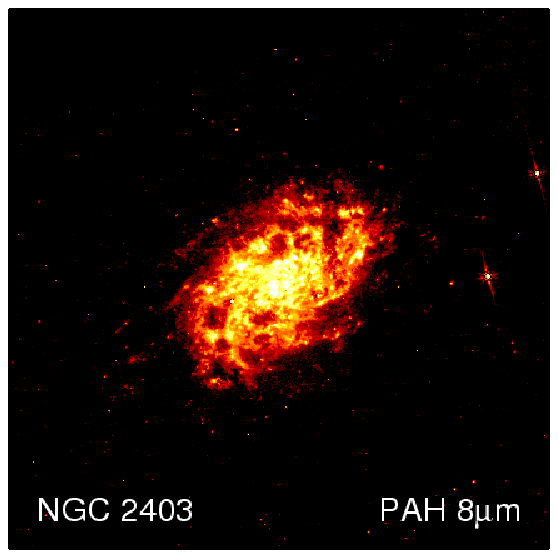, height=50mm}
\epsfig{file=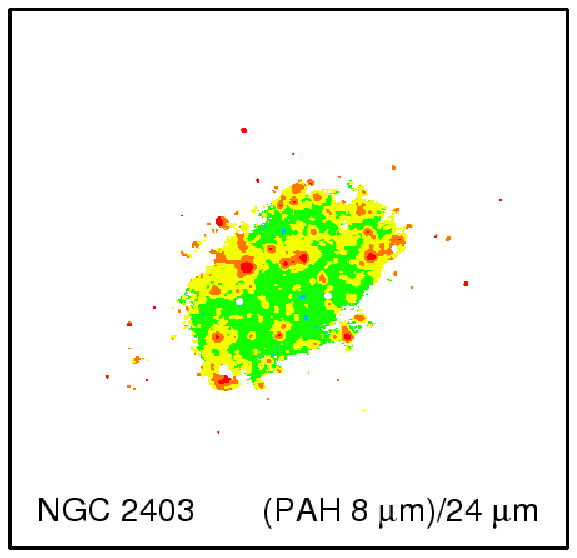, height=50mm}
\epsfig{file=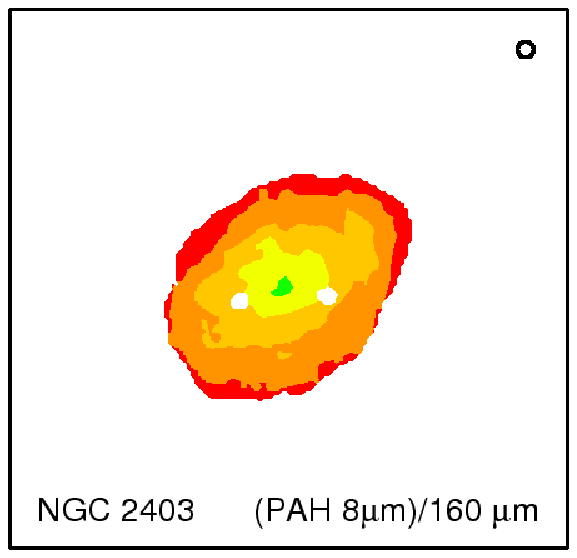, height=50mm}
\epsfig{file=bendog_fig1colourbar1.ps, width=50mm}
\epsfig{file=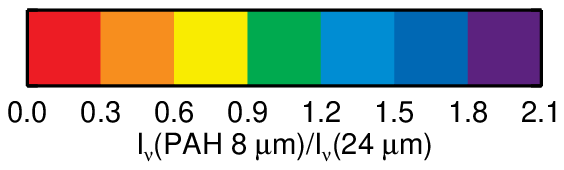, width=50mm}
\epsfig{file=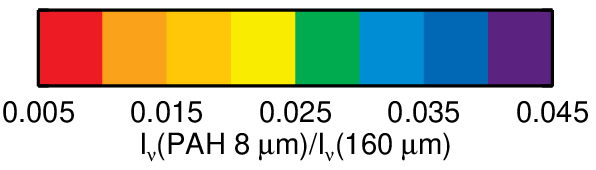, width=50mm}
\end{center}
\caption{Images of the PAH 8~$\mu$m emission at the resolution of the
IRAC data (left column), the (PAH 8~$\mu$m)/24~$\mu$m surface
brightness ratio at the resolution of the 24~$\mu$m data (centre
column), and the (PAH 8~$\mu$m)/160~$\mu$m surface brightness ratio at
the resolution of the 160~$\mu$m data (right column) for the 15
galaxies studied in this analysis.  The three maps for each galaxy are
scaled to the same size.  North is up and east is left in each map.
The stellar continuum has been subtracted from the 8 and 24~$\mu$m
bands to produce these maps.  The contour levels in the ratio maps are
chosen to show structure without showing excessive scatter from noise.
The colour bars on this page give the values of the contour levels.
The red colours in the ratio maps correspond to relatively weak PAH
8~$\mu$m emission, and the blue colours correspond to relatively
strong PAH 8~$\mu$m emission.  White regions in the ratio maps
correspond to regions with low signal-to-noise ratios, regions with
foreground stars, or regions strongly affected by artefacts in the
data.  In the PAH 8~$\mu$m image of NGC 3351, strips of data to the
northeast and southwest of the centre have been strongly affected by
muxbleed and have been masked out in the ratio maps.  The circle in
the top right corner of the (PAH 8~$\mu$m)/160~$\mu$m ratio maps shows
the 38~arcsec FWHM of the 160~$\mu$m beam.}
\label{f_map}
\end{figure*}

\begin{figure*}
\begin{center}
\epsfig{file=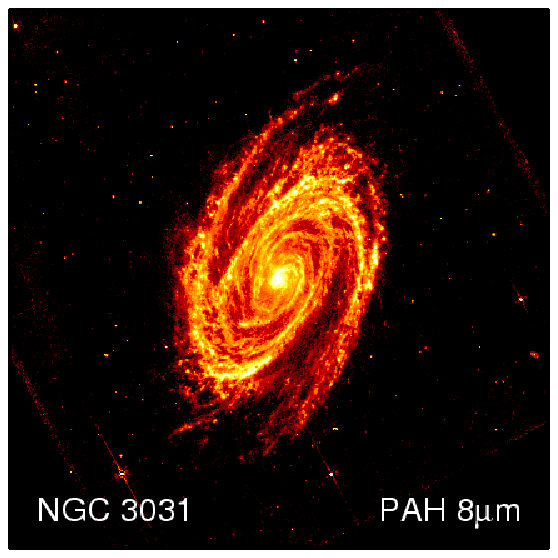, height=50mm}
\epsfig{file=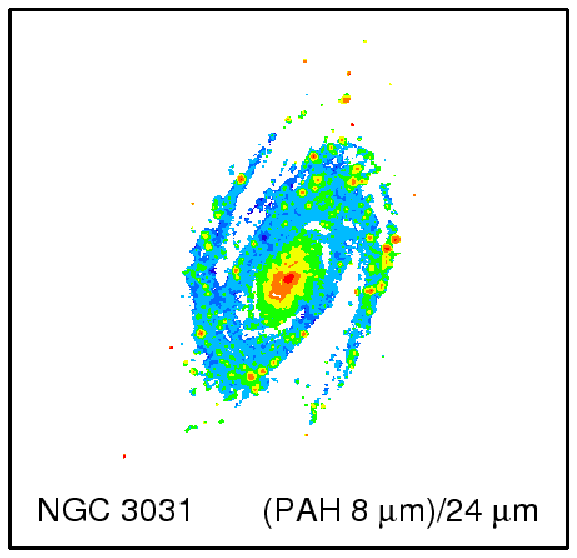, height=50mm}
\epsfig{file=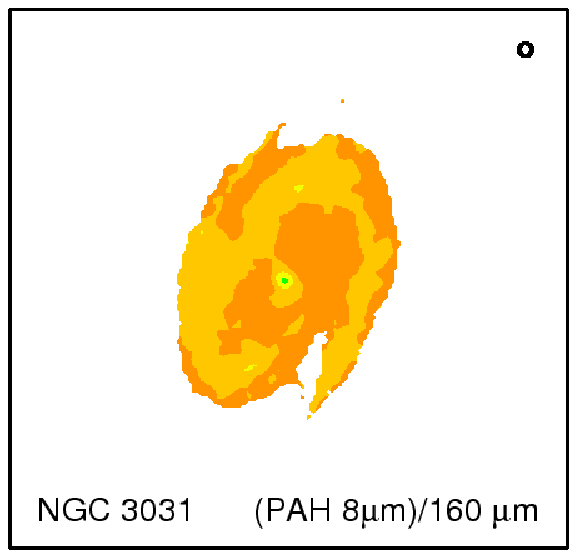, height=50mm}\linebreak
\epsfig{file=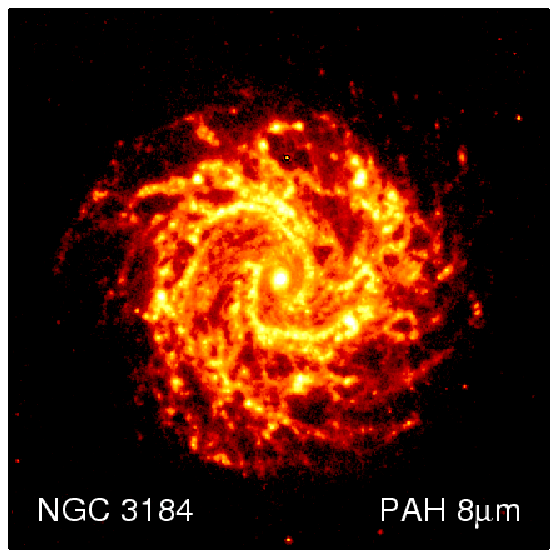, height=50mm}
\epsfig{file=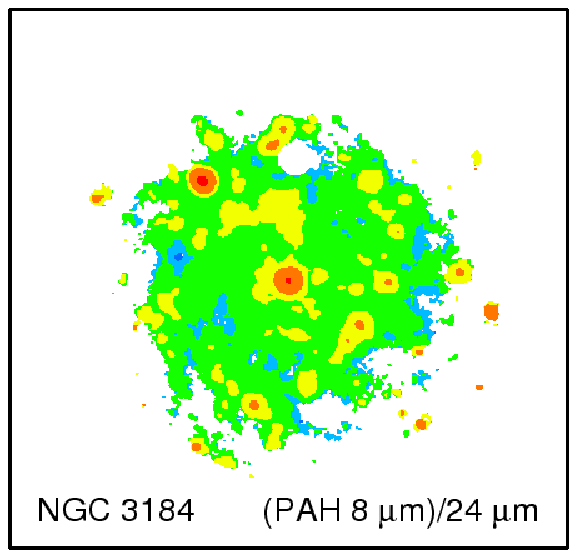, height=50mm}
\epsfig{file=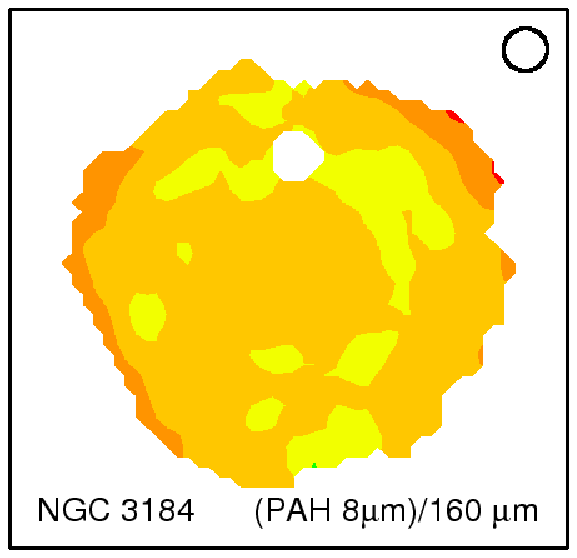, height=50mm}\linebreak
\epsfig{file=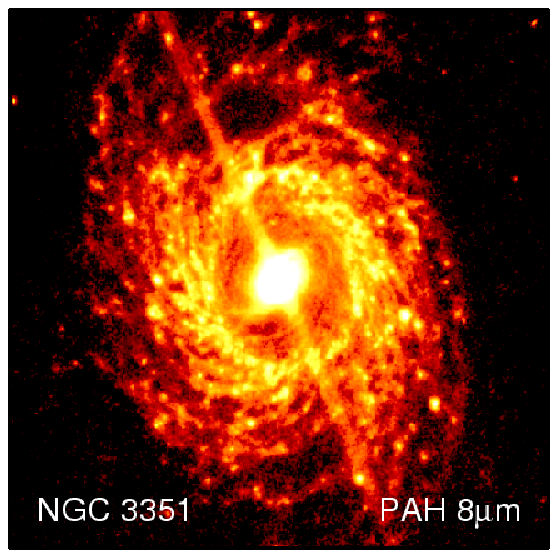, height=50mm}
\epsfig{file=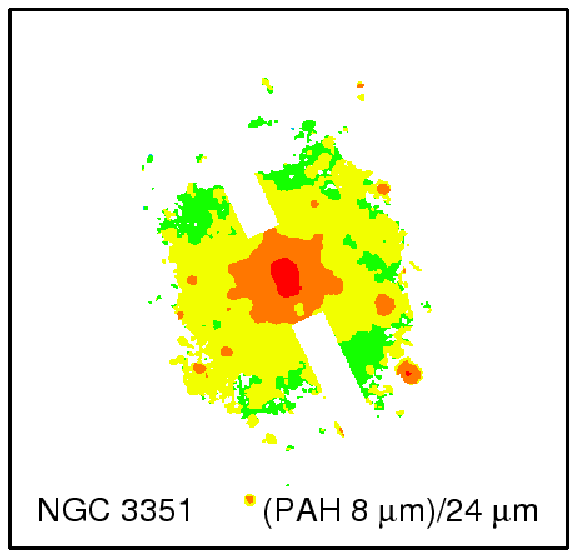, height=50mm}
\epsfig{file=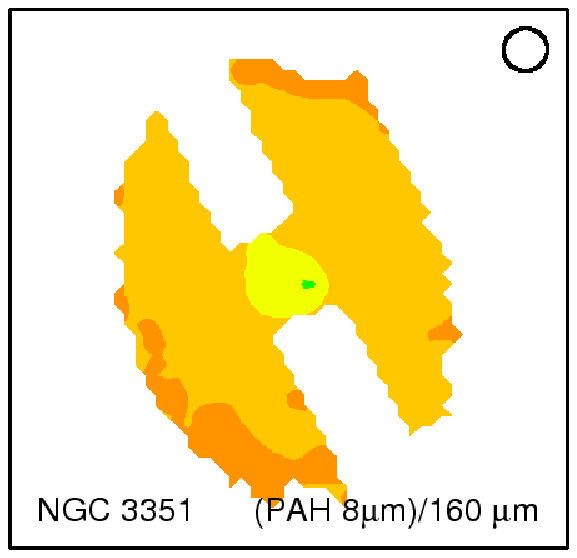, height=50mm}\linebreak
\epsfig{file=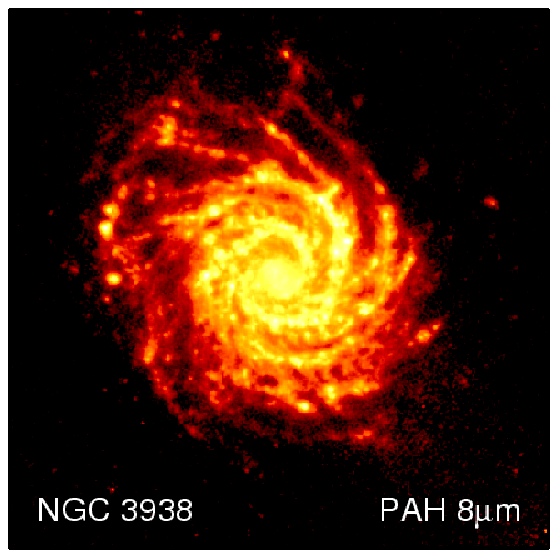, height=50mm}
\epsfig{file=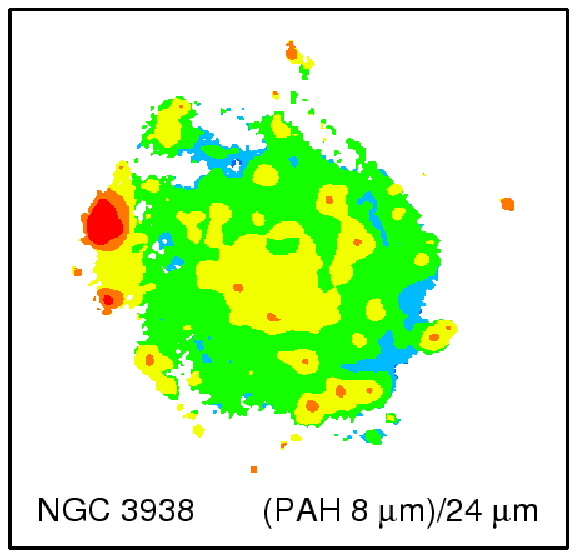, height=50mm}
\epsfig{file=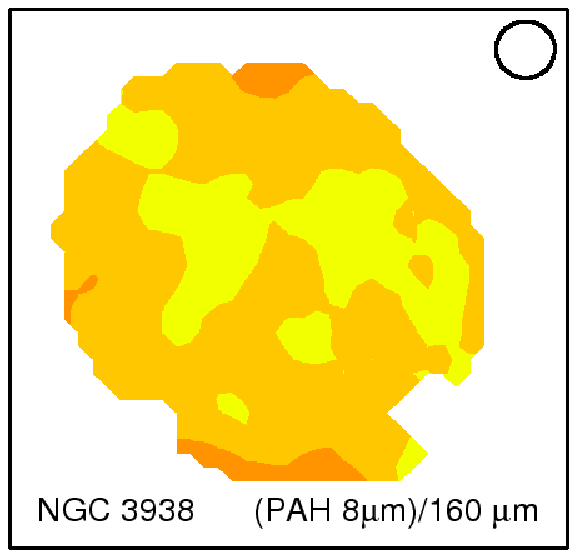, height=50mm}\linebreak
\epsfig{file=bendog_fig1colourbar1.ps, width=50mm}
\epsfig{file=bendog_fig1colourbar2.ps, width=50mm}
\epsfig{file=bendog_fig1colourbar3.ps, width=50mm}
\end{center}
\contcaption{}
\end{figure*}

\begin{figure*}
\begin{center}
\epsfig{file=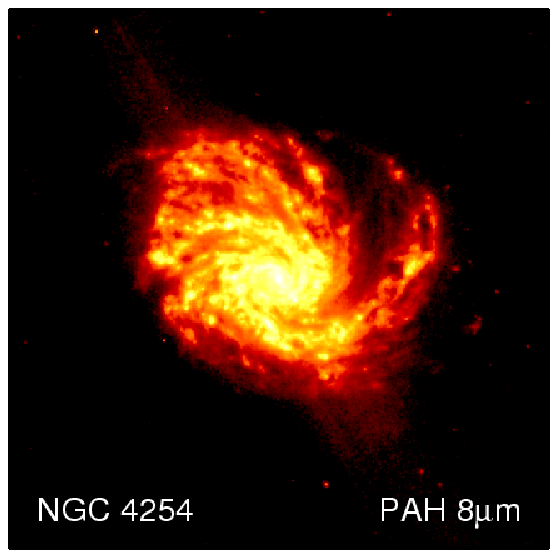, height=50mm}
\epsfig{file=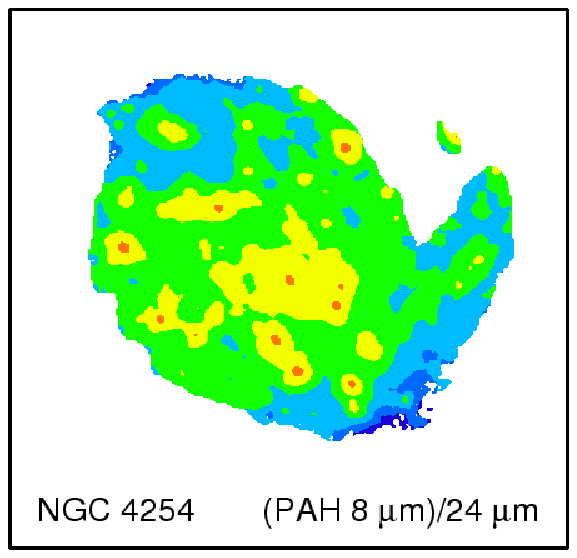, height=50mm}
\epsfig{file=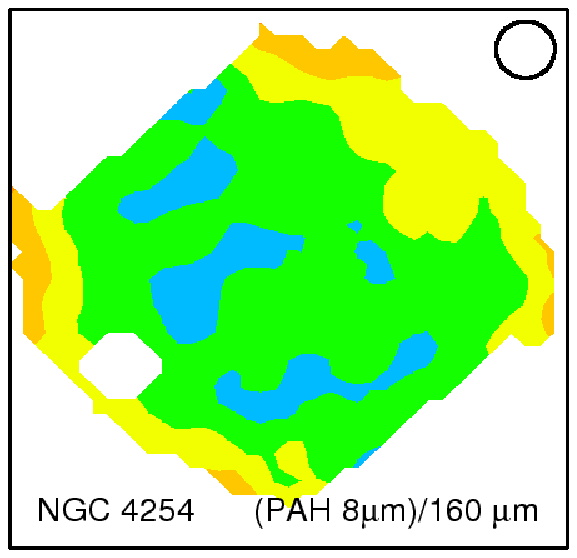, height=50mm}\linebreak
\epsfig{file=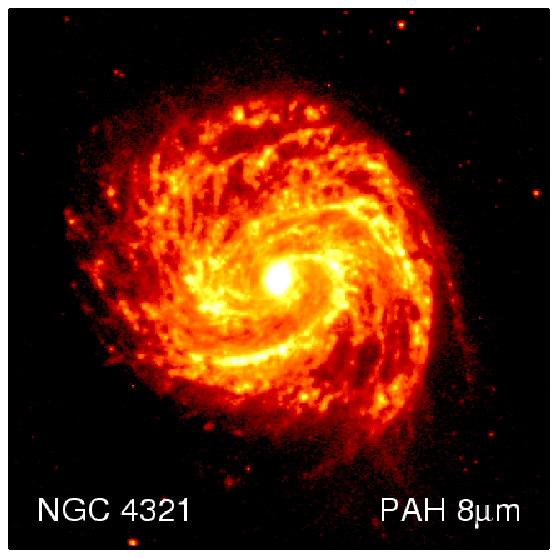, height=50mm}
\epsfig{file=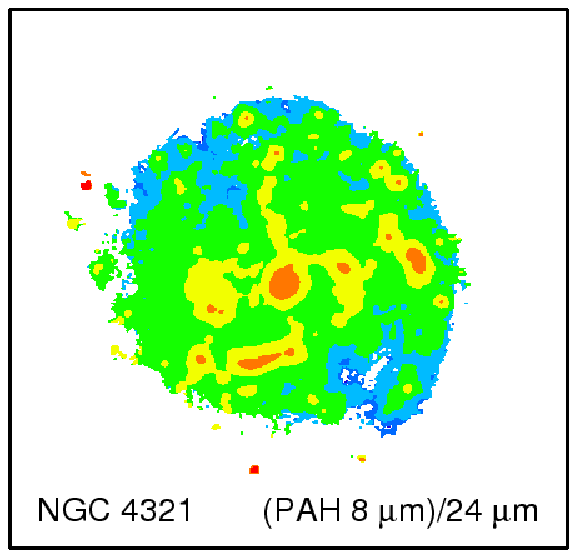, height=50mm}
\epsfig{file=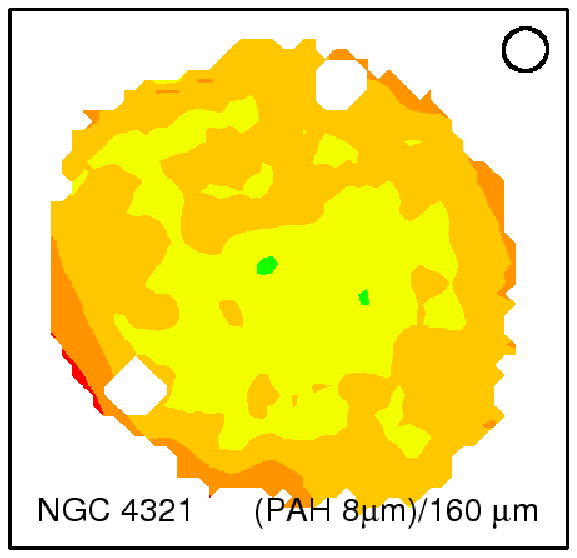, height=50mm}\linebreak
\epsfig{file=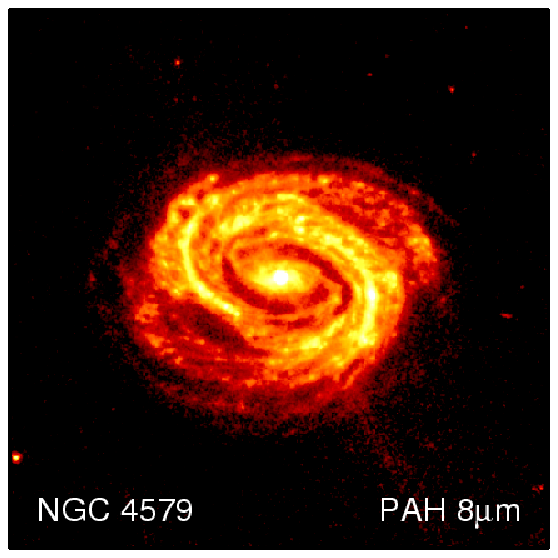, height=50mm}
\epsfig{file=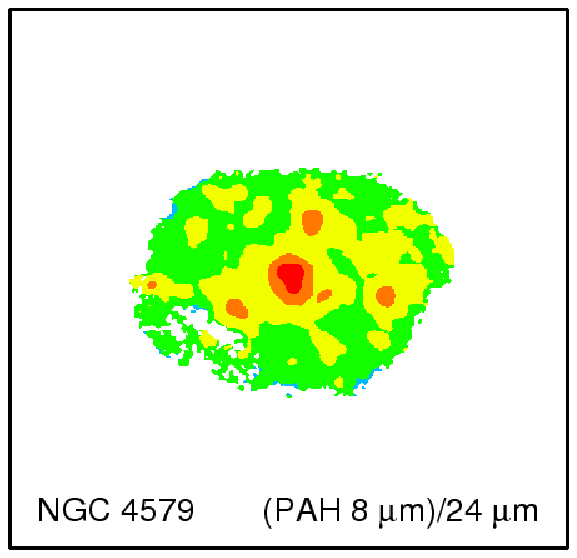, height=50mm}
\epsfig{file=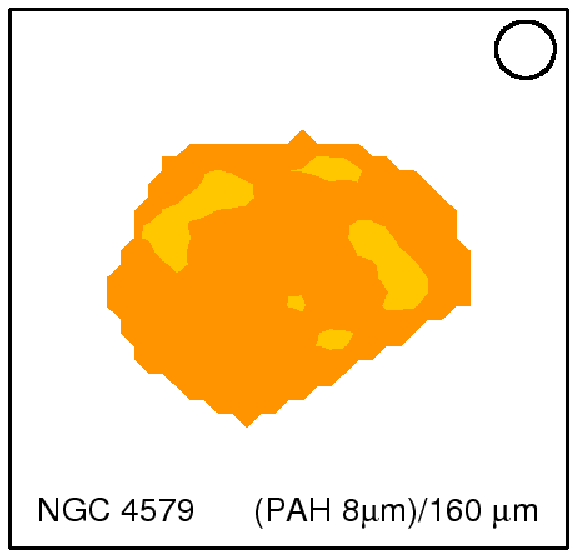, height=50mm}\linebreak
\epsfig{file=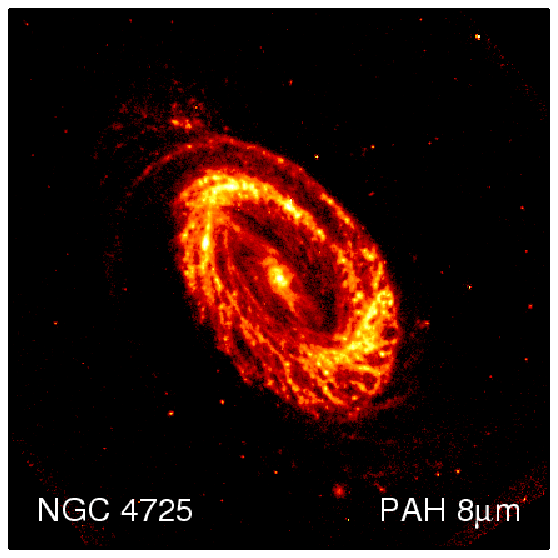, height=50mm}
\epsfig{file=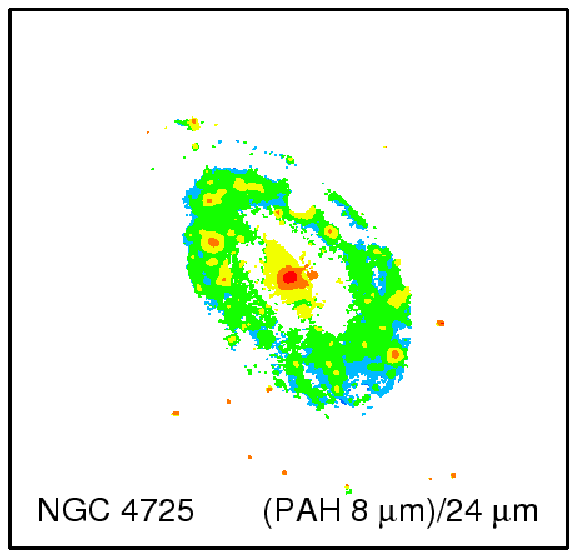, height=50mm}
\epsfig{file=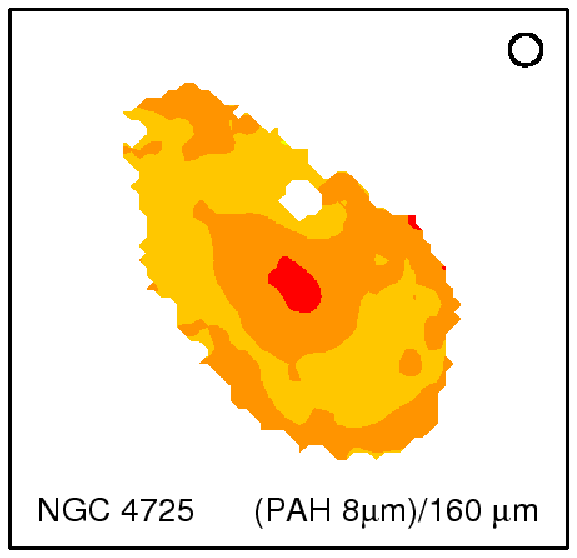, height=50mm}\linebreak
\epsfig{file=bendog_fig1colourbar1.ps, width=50mm}
\epsfig{file=bendog_fig1colourbar2.ps, width=50mm}
\epsfig{file=bendog_fig1colourbar3.ps, width=50mm}
\end{center}
\contcaption{}
\end{figure*}

\begin{figure*}
\begin{center}
\epsfig{file=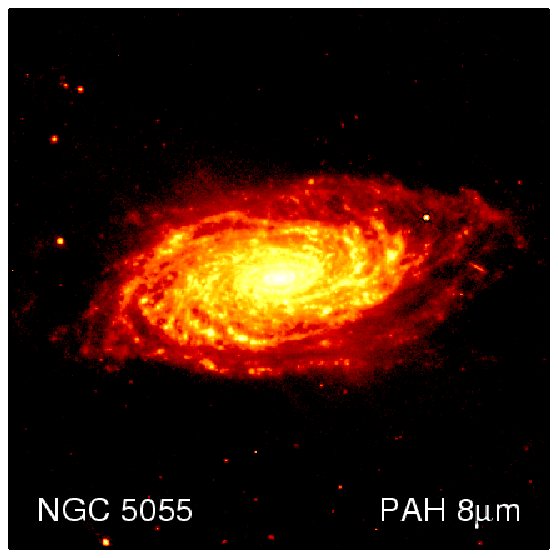, height=50mm}
\epsfig{file=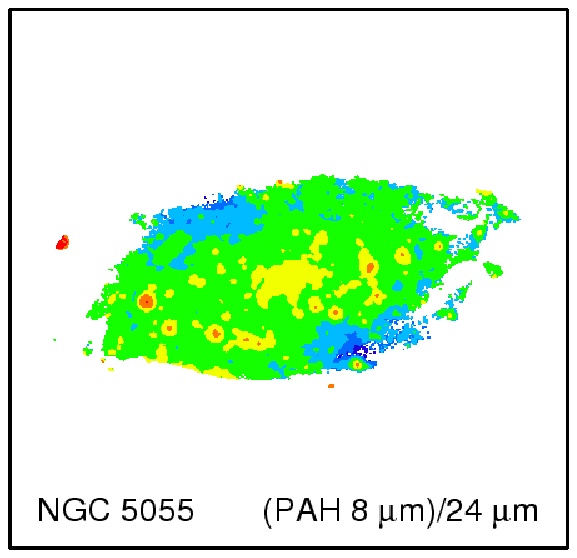, height=50mm}
\epsfig{file=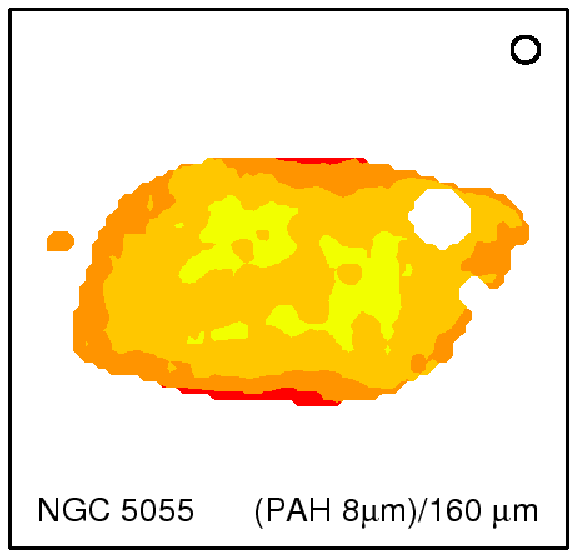, height=50mm}\linebreak
\epsfig{file=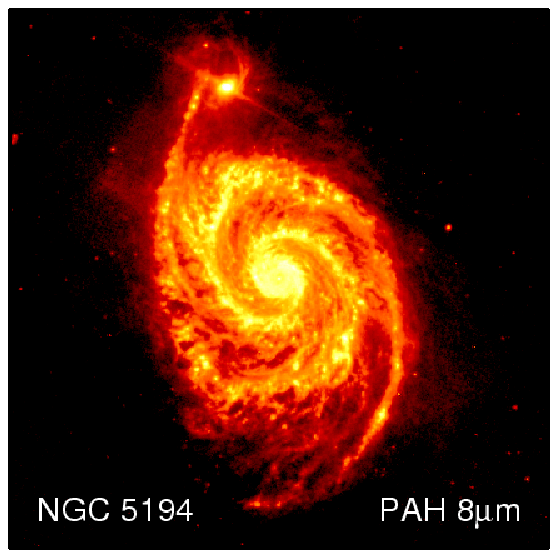, height=50mm}
\epsfig{file=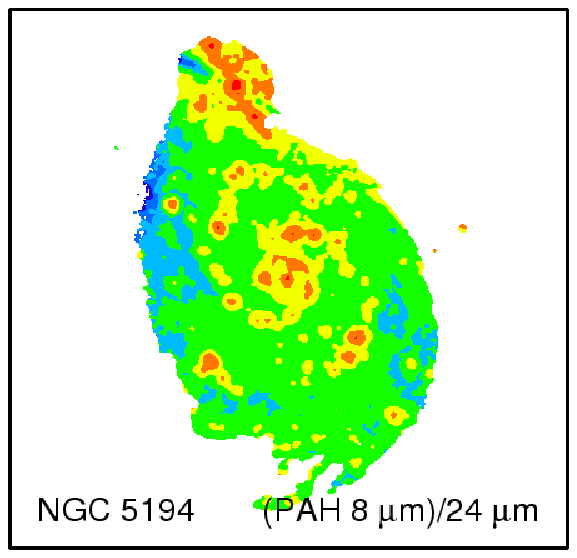, height=50mm}
\epsfig{file=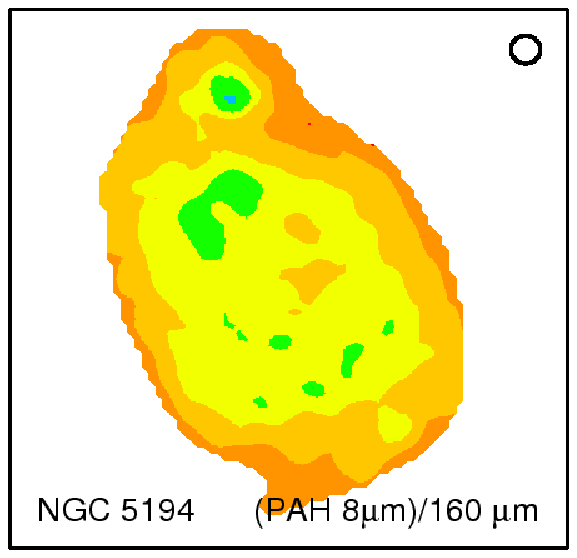, height=50mm}\linebreak
\epsfig{file=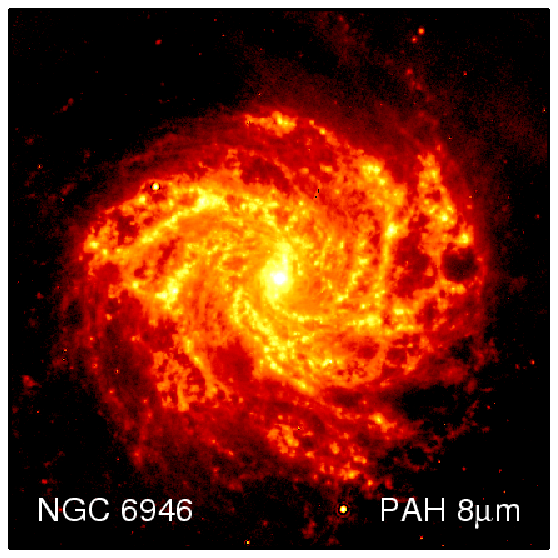, height=50mm}
\epsfig{file=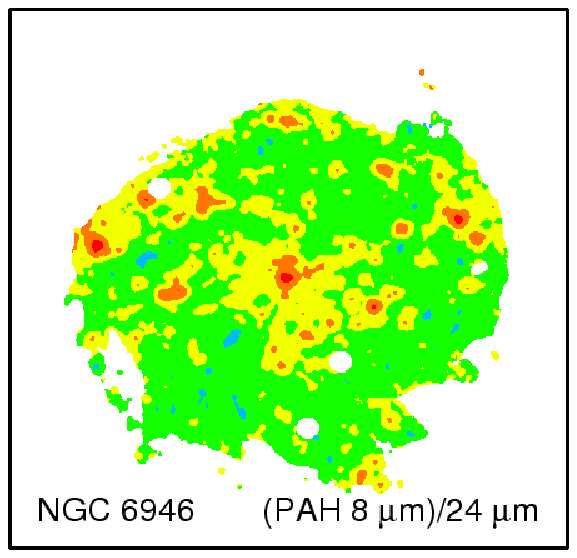, height=50mm}
\epsfig{file=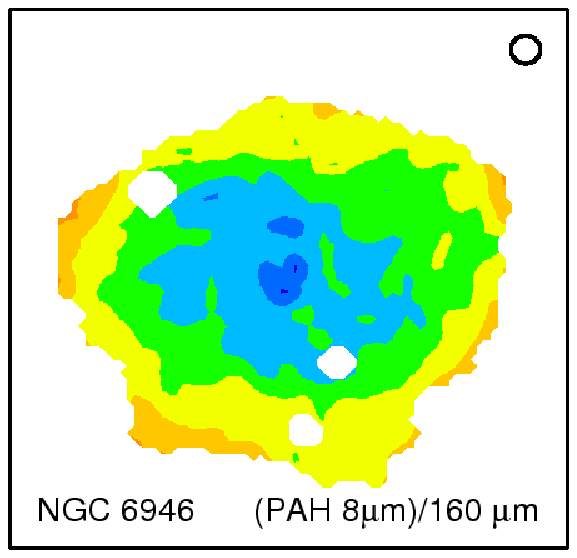, height=50mm}\linebreak
\epsfig{file=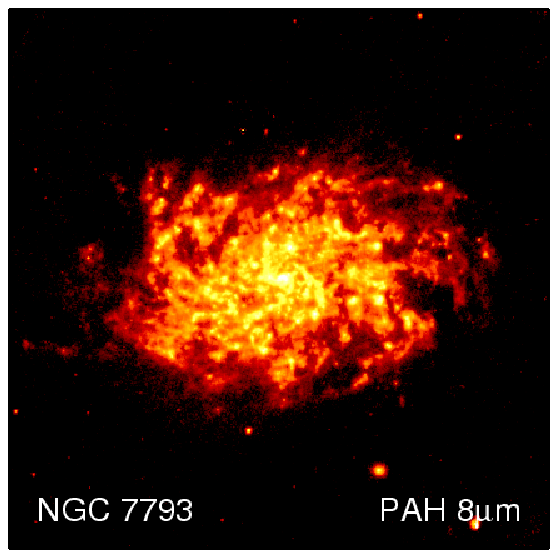, height=50mm}
\epsfig{file=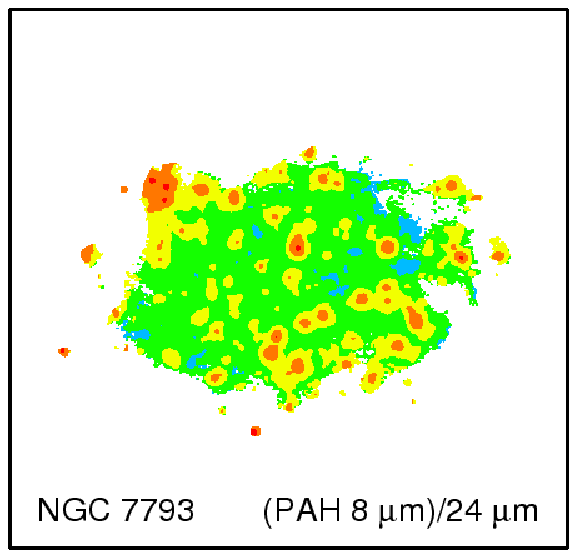, height=50mm}
\epsfig{file=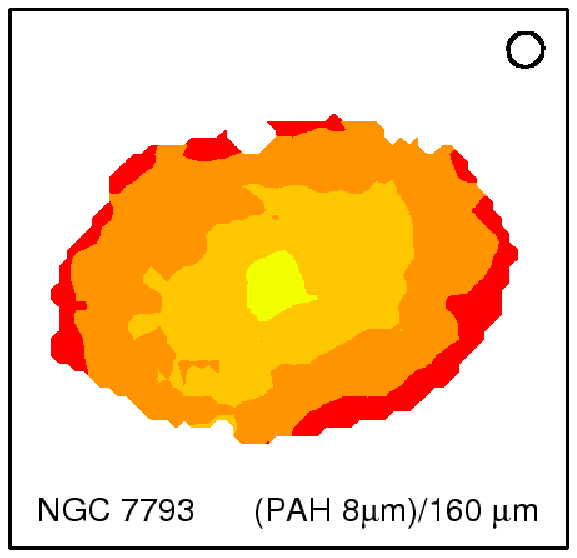, height=50mm}\linebreak
\epsfig{file=bendog_fig1colourbar1.ps, width=50mm}
\epsfig{file=bendog_fig1colourbar2.ps, width=50mm}
\epsfig{file=bendog_fig1colourbar3.ps, width=50mm}
\end{center}
\contcaption{}
\end{figure*}

\section{Comparisons of PAH 8 and 24~$\mu$m emission}
\label{s_comp_pah24}

\begin{figure*}
\epsfig{file=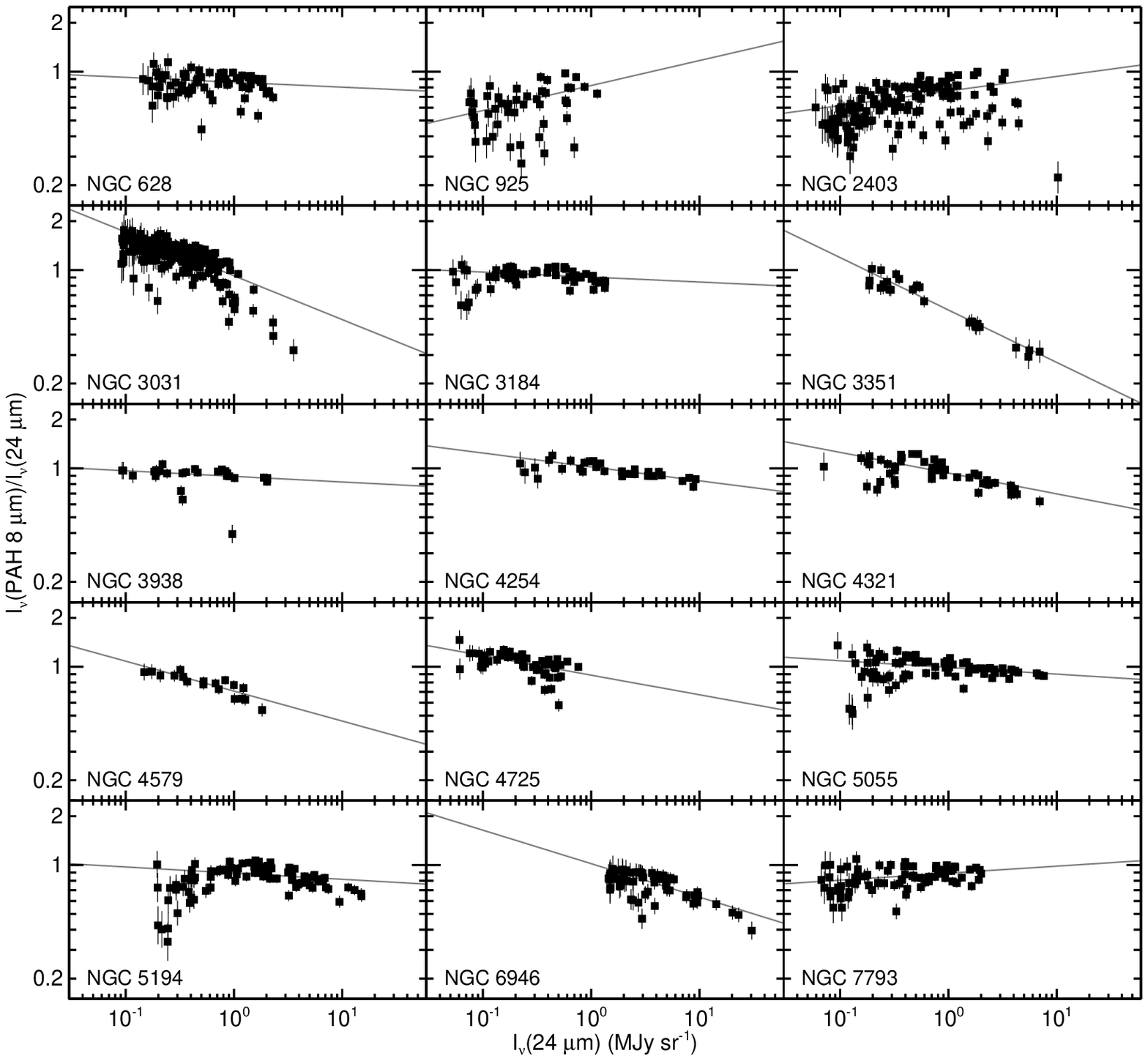}
\caption{Plots of the (PAH 8~$\mu$m)/24~$\mu$m surface brightness
ratios versus the 24~$\mu$m surface brightnesses for the 45~arcsec
square regions measured in these galaxies.  Note that the data are
extracted from images with the same resolution as the 160~$\mu$m data.
This was done so that resolution effects will not be a factor when
comparing the data in this figure to the data in
Figure~\ref{f_pahvs160}.  The grey lines are the best fitting lines
for the relations in each plot; slopes and intrinsic scatters for
these fits are given in Table~\ref{t_pahvs24}.  Note that the
uncertainties in the x- and y-directions are used to weight the data
in the fit.}
\label{f_pahvs24}
\end{figure*}

The panels in the middle columns of Figure~\ref{f_map} show how the
(PAH 8~$\mu$m)/24~$\mu$m surface brightness ratio varies within the
discs of the galaxies.  The resolution of the maps is 6~arcsec, which
allows for seeing fine details within the images.  If the PAH 8~$\mu$m
surface brightness varies linearly with the 24~$\mu$m surface
brightness, then this quantity should be constant across the discs of
these galaxies.  Instead, the (PAH 8~$\mu$m)/24~$\mu$m ratio is seen
to vary significantly.  In particular, the 24~$\mu$m emission peaks
more strongly in point-like regions, which, according to the results
from \citet{cetal07} and \citet{pkbetal07}, are H{\small II} regions.
In contrast, the PAH 8~$\mu$m emission is stronger relative to the
24~$\mu$m emission in the diffuse interstellar regions.  This is most
apparent when comparing the arm and interarm regions of the
grand-design spiral galaxies in this sample, notably NGC~628,
NGC~3031, and NGC~5194.  Some H{\small II} regions well outside the
centres of the galaxies have very low (PAH 8~$\mu$m)/24~$\mu$m ratios
even compared to the rest of the optical discs of the galaxies, as can
be seen with the H{\small II} region in the northeast part of the disc
in NGC~3184 and the H{\small II} region in the east part of the disc
in NGC~3938.

Plots comparing colour to surface brightness are used here and in the
next section to study the correlation between two wave bands.  In a
first-order approximation, the PAH 8, 24, and 160~$\mu$m surface
brightnesses are all correlated with one another for these data.
Hence, colour variations may be difficult to see or measure in plots
directly comparing surface brightnesses in two wave bands.  This is
because the surface brightnesses vary by factors of $\sim100$ but the
ratios of the surface brightnesses vary by less than a factor of 10.
However, plots comparing surface brightness ratios to surface
brightnesses can more clearly show deviations in the relation between
two wave bands, including systematic variations in colour related to
surface brightness.  If a one-to-one correspondence exists between the
PAH and 24~$\mu$m surface brightnesses, then the slopes of the best
fit lines in these plots would be close to 0, and the scatter in these
plots would be small.  Such plots would be biased towards producing
relations with slopes of -1 in log-log space if the two wave bands
were randomly distributed.  However, since the flux densities in all
wave bands we studied here are approximately proportional to each
other, such biases will not be present.  To measure the physical
(intrinsic) scatter around the best fit line, we will subtract the sum
of the square of the measurement uncertainties from the sum of the
square of the residuals from the best fit line.  This is given by
\begin{equation}
s^2 = \sum{(y_i-a-bx_i)^2} - \sum{(\sigma_{xi}^2+b^2\sigma_{yi}^2)}
\label{e_scatter}
\end{equation}
where $x$ and $y$ are abscissa and ordinate values with corresponding
uncertainties $\sigma_x$ and $\sigma_y$ and $a$ and $b$ are the
y-intercept and slope of of the best fit line.  A value of 0 is
reported if the result from Equation~\ref{e_scatter} is negative, as
this would indicate that all of the scatter around the best fit line
could be accounted for by the measurement uncertainties.

\begin{table}
\begin{center}
\renewcommand{\thefootnote}{\alph{footnote}}
\caption{Slopes and Intrinsic Scatter for the Best Fit Line Describing (PAH
8)~/~24~$\mu$m Surface Brightness Ratio versus 24~$\mu$m Surface Brightness
Density$^a$ \label{t_pahvs24}}
\begin{tabular}{@{}lcc@{}}
\hline
Name &        Slope &                 Intrinsic\\
&             &                       Scatter\\
\hline
NGC 628 &     $-0.030 \pm 0.015$ &    0.20 \\
NGC 925 &     $0.15 \pm 0.02$  &      0.69 \\
NGC 2403 &    $0.090 \pm 0.008$ &     1.91 \\
NGC 3031 &    $-0.270 \pm 0.011$ &    0.93 \\
NGC 3184 &    $-0.030 \pm 0.011$ &    0.13 \\
NGC 3351 &    $-0.32 \pm 0.02$ &      0 \\
NGC 3938 &    $-0.035 \pm 0.015$ &    0.14 \\
NGC 4254 &    $-0.085 \pm 0.013$ &    0.0011 \\
NGC 4321 &    $-0.128 \pm 0.010$ &    0.14 \\
NGC 4579 &    $-0.18 \pm 0.03$ &      0 \\
NGC 4725 &    $-0.121 \pm 0.017$ &    0.12 \\
NGC 5055 &    $-0.041 \pm 0.007$ &    0.43 \\
NGC 5194 &    $-0.038 \pm 0.008$ &    0.89 \\
NGC 6946 &    $-0.21 \pm 0.02$ &      0.069 \\
NGC 7793 &    $0.043 \pm 0.011$ &     0.13 \\
\hline
\end{tabular}
\end{center}
$^a$ These slopes and intrinsic scatter measurements are for the lines fit
     to the data in Figure~\ref{f_pahvs24}.  The data are measured
     within 45~arcsec bins in images with the same resolution as the
     160~$\mu$m images. Intrinsic scatter measurements of 0 are reported
     when the value calculated with Equation~\ref{e_scatter} is
     negative; this indicates that measurement uncertainties can
     account for all of the scatter in the relation.
\end{table}
\renewcommand{\thefootnote}{\arabic{footnote}}

Figure~\ref{f_pahvs24} shows how the (PAH 8~$\mu$m)/24~$\mu$m ratio
varies with 24~$\mu$m surface brightness among the 45~arcsec square
regions described in Section~\ref{s_dataprep}.  Note that the
resolution of the data used in this figure is matched to the
38~arcsec resolution of the 160~$\mu$m images so that the results from
these figures can be more easily compared to the results in
Section~\ref{s_comp_pah160}.  The slopes and intrinsic scatter for the
best fit lines are given in Table~\ref{t_pahvs24}.  The best fitting
lines are determined using uncertainties in both the x- and
y-directions to weight the data, so the fits are strongly weighted
towards high surface brightness regions.

\begin{figure*}
\epsfig{file=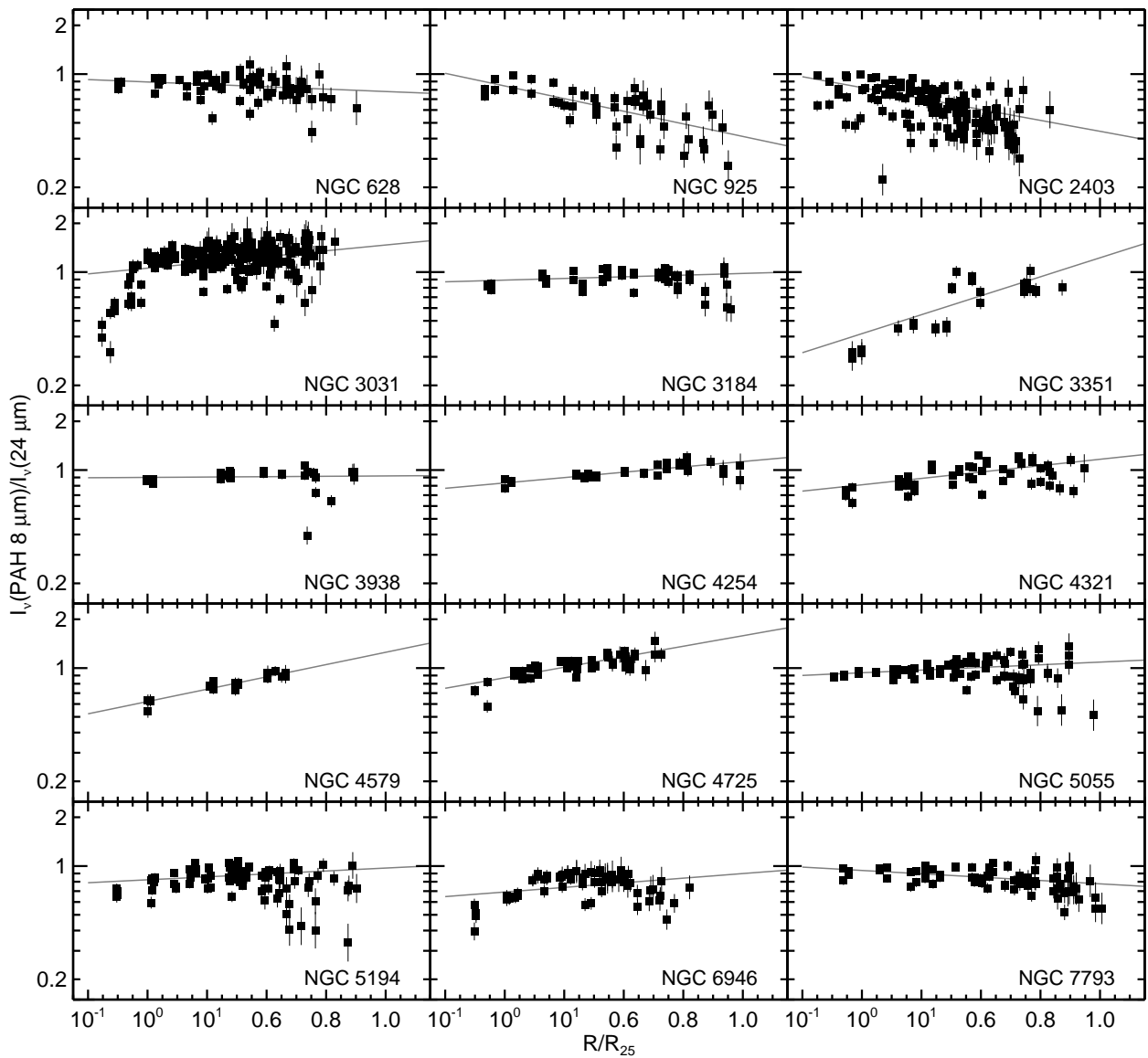}
\caption{Plots of the (PAH 8~$\mu$m)/24~$\mu$m surface brightness
ratios versus deprojected galactocentric radii for the 45~arcsec
square regions measured in these galaxies.  Note that the data are
extracted from images with the same resolution as the 160~$\mu$m data.
The radii are normalized by the radius of the D$_{25}$ isophote given
by \citet{ddcbpf91}.  The grey lines are the best fitting lines for
the relations in each plot; slopes and intrinsic scatter for these
fits are given in Table~\ref{t_pah160vsdist}.  Only the uncertainties
in the (PAH 8~$\mu$m)/160~$\mu$m ratio are used to weight the data in
the fit; the uncertainties in the radius are assumed to be negligible
in this analysis.}
\label{f_pah24vsdist}
\end{figure*}

The slopes of the best fit lines in Figure~\ref{f_pahvs24} are
generally not statistically equivalent to 0, which demonstrates that
the (PAH 8~$\mu$m)/24~$\mu$m ratio varies with surface brightness.  In
many galaxies (most notably NGC~3031, NGC~3351, and NGC 6946), the
(PAH 8~$\mu$m)/24~$\mu$m ratio decreases notably as the 24~$\mu$m
surface brightness increases.  These tend to be galaxies with
infrared-bright point-like nuclei.  In a few other galaxies (NGC~2403
and NGC~7793, for example), the (PAH 8~$\mu$m)/24~$\mu$m ratio
increases as the 24~$\mu$m surface brightness increases.  The data for
some galaxies also show a broad scatter, particularly for galaxies
without infrared-bright nuclei.  For a given 24~$\mu$m surface
brightness, the (PAH~8~$\mu$m)/24~$\mu$m ratio varies by over a factor
of 2 in high signal-to-noise regions in NGC~925 and NGC~2403.
Although both of these galaxies are fit with linear relations in
Figure~\ref{f_pahvs24}, the broad scatter indicates that such a fit is
unrealistic, so the (PAH 8) and 24~$\mu$m emission must be only weakly
associated with each other.  Also, some data points fall well
below the best fit lines in Figure~\ref{f_pahvs24}, particularly in
the plots for NGC~2403 and NGC~3938.  These regions correspond to very
infrared-bright star-forming regions visible in Figure~\ref{f_map}. 

Figure~\ref{f_map} does not reveal the presence of any obvious
dependence of the (PAH 8~$\mu$m)/24~$\mu$m on radius, but it is still
useful to measure such gradients for comparison with abundance
gradients, especially since it has been shown that the (PAH
8~$\mu$m)/24~$\mu$m ratio varies with metallicity \citep{eetal05,
detal05, ddbetal07, cetal07, eetal08}.  The relation between the (PAH
8~$\mu$m)/24~$\mu$m ratio and radius for the 45~arcsec square regions
described in Section~\ref{s_dataprep} is shown in
Figure~\ref{f_pah24vsdist}, with slopes for the best fit lines and
intrinsic scatter given in Table~\ref{t_pah24vsdist}.  As in
Figure~\ref{f_pahvs24}, the slopes vary significantly among the
galaxies in this sample.  Some galaxies with infrared-bright nuclei,
such as NGC~3351 and NGC~4579, have large positive radial gradients in
the (PAH 8~$\mu$m)/24~$\mu$m ratio, and in NGC~3031 and NGC~6946, the
infrared-bright nuclei fall below the relation between the (PAH
8~$\mu$m)/24~$\mu$m ratio and radius.  In a few other galaxies (most
notably NGC~925 and NGC~2403), the radial gradient in the (PAH
8~$\mu$m)/24~$\mu$m ratio is negative.  As in the relation between the
(PAH 8~$\mu$m)/24~$\mu$m ratio versus 24~$\mu$m surface brightness,
significant scatter is seen in some relations between the (PAH
8~$\mu$m)/24~$\mu$m ratio and radius, and some infrared-bright
regions fall well below the best fit lines, as can be seen most
clearly in NGC~2403 and NGC~3938.  Further discussion concerning how
these gradients might be related to radial gradients in abundances is
presented in Section~\ref{s_discuss_pah24}.

\begin{table}
\begin{center}
\renewcommand{\thefootnote}{\alph{footnote}}
\caption{Slopes and Intrinsic Scatter for the Best Fit Line Describing (PAH
8)~/~24~$\mu$m Surface Brightness Ratio versus Radius$^a$
\label{t_pah24vsdist}}
\begin{tabular}{@{}lcc@{}}
\hline
Name &        Slope &                 Intrinsic\\
&             &                       Scatter\\
\hline
NGC 628 &     $-0.07 \pm 0.02$ &      0.16 \\
NGC 925 &     $-0.39 \pm 0.03$  &     0.23 \\
NGC 2403 &    $-0.34 \pm 0.02$ &      1.43 \\
NGC 3031 &    $0.178 \pm 0.016$ &     1.64 \\
NGC 3184 &    $0.051 \pm 0.019$ &     0.13 \\
NGC 3351 &    $0.59 \pm 0.05$ &       0.18 \\
NGC 3938 &    $0.01 \pm 0.03$ &       0.15 \\
NGC 4254 &    $0.17 \pm 0.02$ &       0 \\
NGC 4321 &    $0.20 \pm 0.02$ &       0.17 \\
NGC 4579 &    $0.38 \pm 0.05$ &       0 \\
NGC 4725 &    $0.32 \pm 0.03$ &       0.040 \\
NGC 5055 &    $0.082 \pm 0.017$ &     0.42 \\
NGC 5194 &    $0.09 \pm 0.02$ &       0.85 \\
NGC 6946 &    $0.14 \pm 0.03$ &       0.23 \\
NGC 7793 &    $-0.103 \pm 0.018$ &    0.091 \\
\hline
\end{tabular}
\end{center}
$^a$ These slopes and intrinsic scatter measurements are for the lines fit
     to the data in Figure~\ref{f_pah24vsdist}.  The data are measured
     within 45~arcsec bins in images with the same resolution as the
     160~$\mu$m images. Intrinsic scatter measurements of 0 are reported
     when the value calculated with Equation~\ref{e_scatter} is
     negative; this indicates that measurement uncertainties can
     account for all of the scatter in the relation.
\end{table}
\renewcommand{\thefootnote}{\arabic{footnote}}

Overall, these data demonstrate that the relation between PAH 8 and
24~$\mu$m emission on spatial scales smaller than $\sim2$~kpc may
exhibit a significant amount of scatter.  Moreover, the relation
between the (PAH 8~$\mu$m)/24~$\mu$m ratio and the 24~$\mu$m surface
brightness does not vary in a way that is easily predictable within
these galaxies.

\section{Comparisons of PAH 8 and 160~$\mu$m emission}
\label{s_comp_pah160}

The panels in the right columns of Figure~\ref{f_map} show how the
(PAH 8~$\mu$m)/160~$\mu$m surface brightness ratio varies within the
sample galaxies.  The resolution of the maps is 38~arcsec.  As with
the (PAH 8~$\mu$m)/24~$\mu$m ratio, the (PAH 8~$\mu$m)/160~$\mu$m
ratio does vary across the discs of these galaxies, although the
variations are notably different.  The (PAH 8~$\mu$m)/160~$\mu$m ratio
generally appears to increase as the 160~$\mu$m surface brightness
increases.  Moreover, the (PAH 8~$\mu$m)/160~$\mu$m ratio also appears
enhanced in large scale structures within the discs of these galaxies,
such as the spiral arms in NGC~3031 and NGC~6946.  In the centres of a
few galaxies, the (PAH 8~$\mu$m)/160~$\mu$m ratio appears to decrease
slightly.  This is most apparent in NGC~3184, NGC~4725, and NGC~5055.
The disc of NGC~4725 contains a ring, so the apparent decrease in the
central (PAH 8~$\mu$m)/160~$\mu$m ratio could partly be related to an
enhancement of the (PAH 8~$\mu$m)/160~$\mu$m ratio in the ring.

\begin{figure*}
\epsfig{file=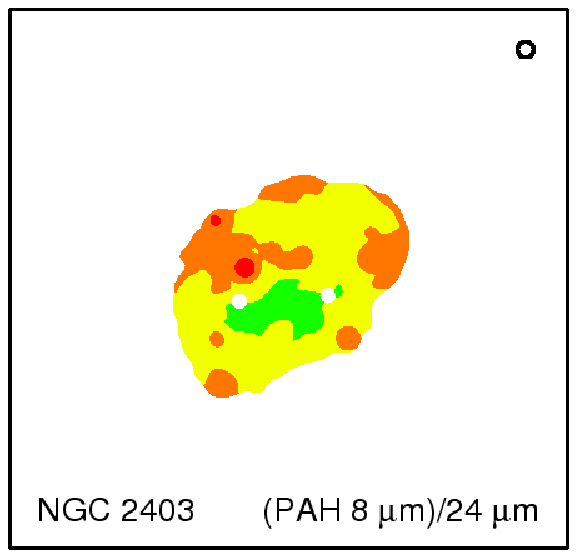, height=58mm}
\epsfig{file=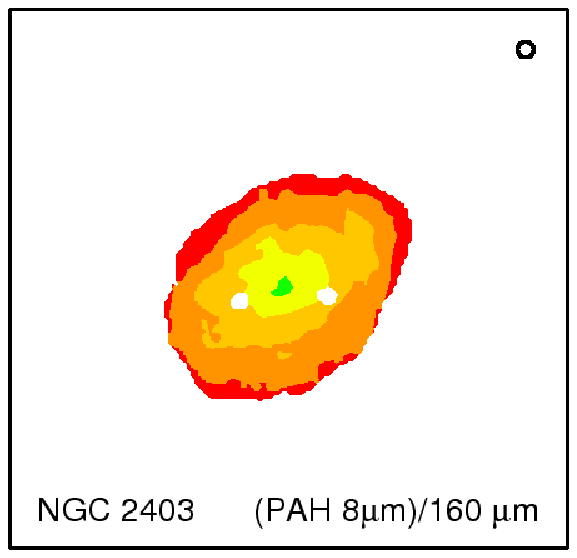, height=58mm} \linebreak
\epsfig{file=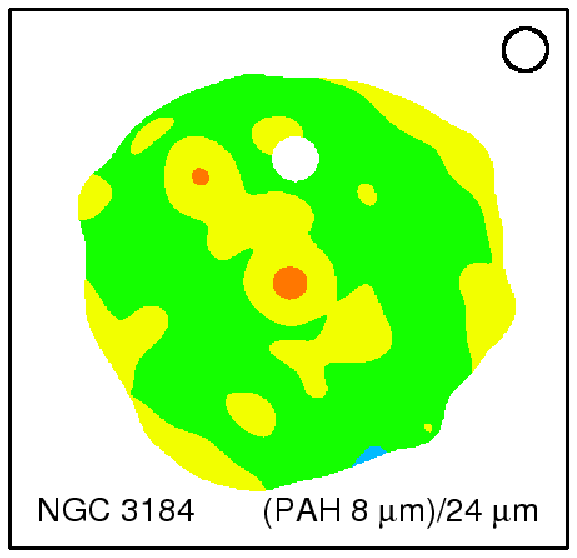, height=58mm}
\epsfig{file=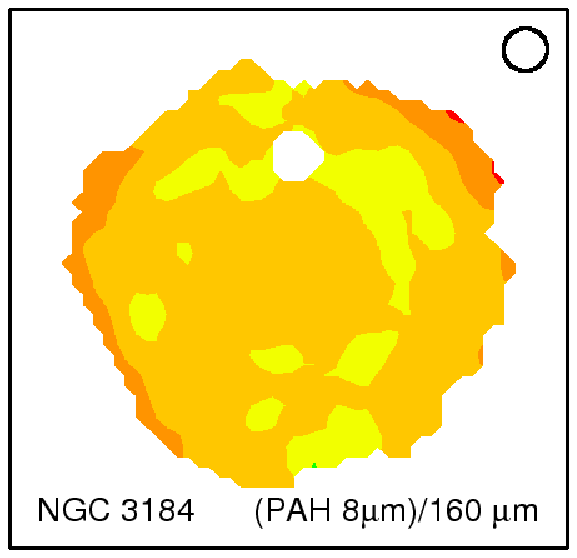, height=58mm} \linebreak
\epsfig{file=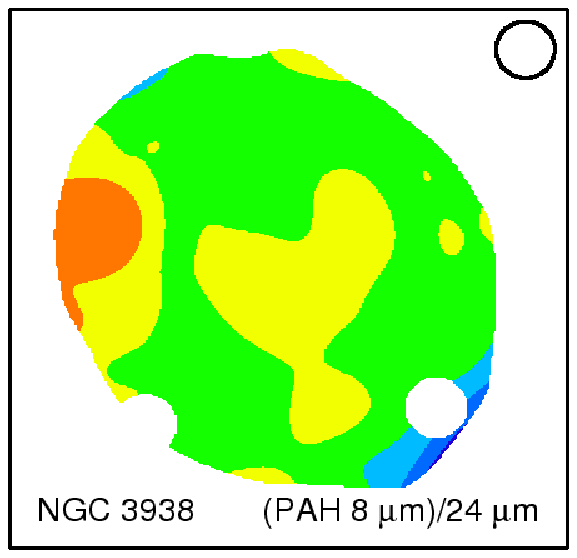, height=58mm}
\epsfig{file=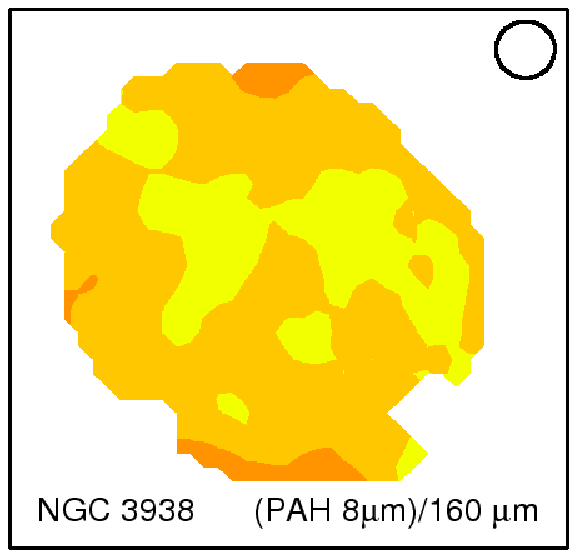, height=58mm} \linebreak
\epsfig{file=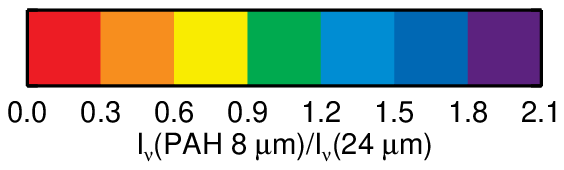, width=58mm}
\epsfig{file=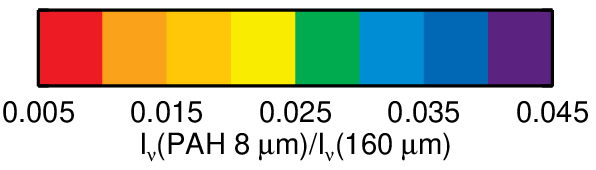, width=58mm}
\caption{Images of the (PAH 8~$\mu$m)/24~$\mu$m surface brightness
ratio at the resolution of the 160~$\mu$m data (left column), and the
(PAH 8~$\mu$m)/160~$\mu$m surface brightness ratio at the resolution
of the 160~$\mu$m data (right column) for three galaxies with
extranuclear star-forming regions that are exceptionally bright at
24~$\mu$m.  These maps demonstrate how the lower resolution affects
the (PAH 8~$\mu$m)/24~$\mu$m ratio maps and how the maps still differ
in comparison to the (PAH 8~$\mu$m)/160~$\mu$m ratio maps.  The circle
in the top right corner of each map shows the 38~arcsec FWHM of the
160~$\mu$m beam.  See Figure~\ref{f_map} for additional information.}
\label{f_map_conv160comp}
\end{figure*}

The maps also show that the (PAH 8~$\mu$m)/160~$\mu$m ratio does not
necessarily peak in the extranuclear star-forming regions where local
minima in the (PAH 8~$\mu$m)/24~$\mu$m ratio are found.  This can be
seen most dramatically in the images of NGC~2403, NGC~3184 and
NGC~3938.  To show that the differences between the (PAH
8~$\mu$m)/24~$\mu$m and (PAH 8~$\mu$m)/160~$\mu$m ratio maps is not a
result of resolution effects for these three galaxies, we show both
ratio maps for data at a resolution of 38~arcsec in
Figure~\ref{f_map_conv160comp}.  Even at this resolution, the
infrared-bright H{\small II} regions located in the northeast part of
the disc near the nucleus of NGC~2403, in the northeast part of the
disc in NGC~3184, and in the east part of the disc in NGC~3938 are
still identifiable in the (PAH 8~$\mu$m)/24~$\mu$m ratio maps.
However, these three infrared-bright H{\small II} regions are
indistinguishable from the nearby dust emission in the (PAH
8~$\mu$m)/160~$\mu$m ratio maps.  Nonetheless, keep in mind that
enhancements in the (PAH 8~$\mu$m)/160~$\mu$m ratios are still visible
in the large scale structures such as the spiral arms in NGC~3031 and
NGC~6946.

\begin{figure*}
\epsfig{file=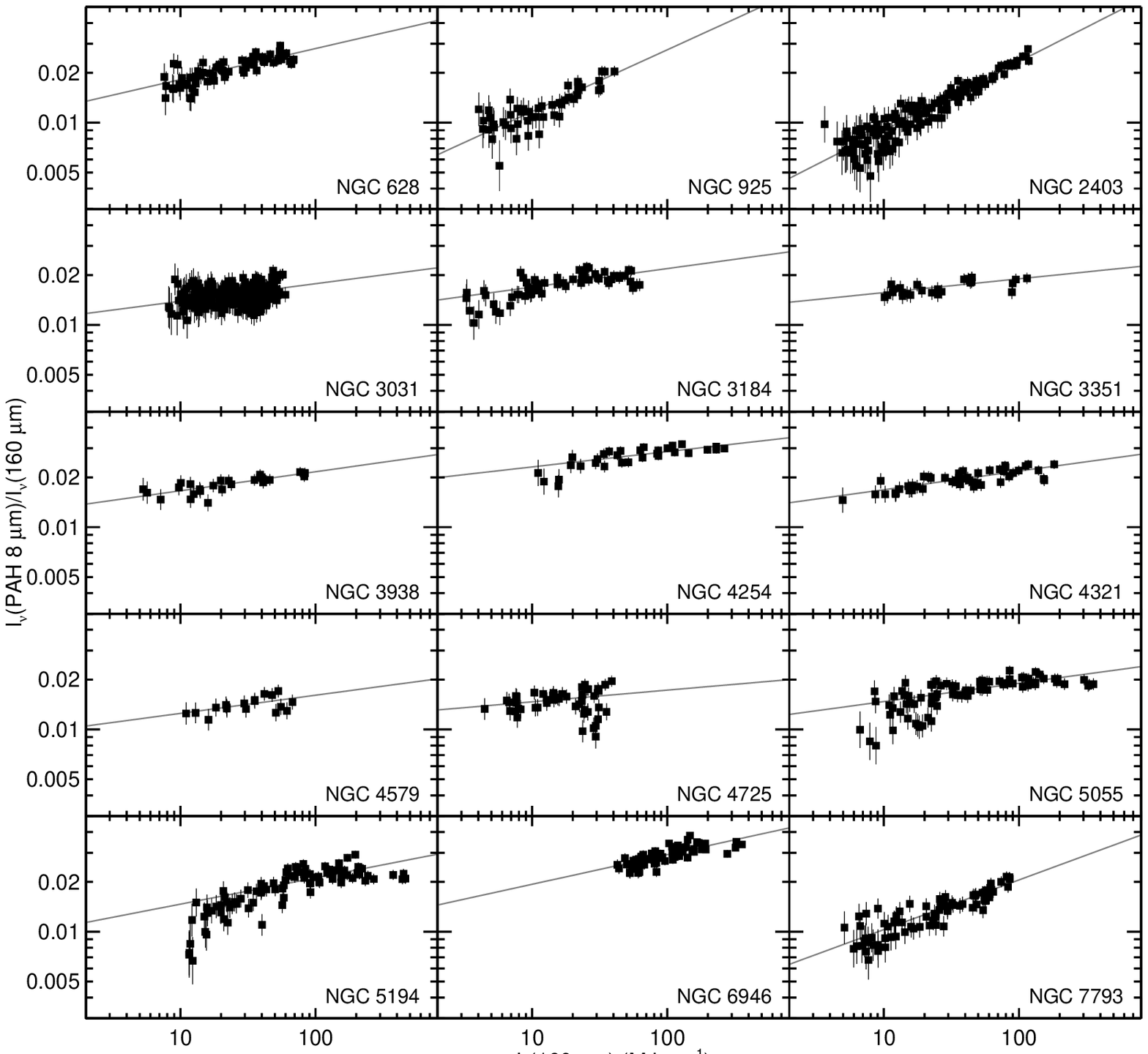}
\caption{Plots of the (PAH 8~$\mu$m)/160~$\mu$m surface brightness
ratios versus the 160~$\mu$m surface brightnesses for the 45~arcsec
square regions measured in these galaxies.  The grey lines are the
best fitting lines for the relations in each plot; slopes, intrinsic
scatter, and correlation coefficients for these fits are given in
Table~\ref{t_pahvs160}.  Note that the uncertainties in the x- and
y-directions are used to weight the data in the fit.}
\label{f_pahvs160}
\end{figure*}

\begin{table}
\begin{center}
\renewcommand{\thefootnote}{\alph{footnote}}
\caption{Results for the Best Fit Line Describing (PAH
  8)~/~160~$\mu$m Surface Brightness Ratio versus 160~$\mu$m Surface
  Brightness$^a$ \label{t_pahvs160}}
\begin{tabular}{@{}lccc@{}}
\hline
Name &        Slope &                Intrinsic &        Spearman's Rank \\
&             &                      Scatter &          Correlation \\
&             &                      &                  Coefficient$^b$ \\
\hline
NGC 628 &     $0.188 \pm 0.018$ &    0 &                0.85 \\
NGC 925 &     $0.37 \pm 0.03$ &      0 &                0.79 \\
NGC 2403 &    $0.415 \pm 0.012$ &    0 &                0.93 \\
NGC 3031 &    $0.11 \pm 0.02$ &      0 &                0.27 \\
NGC 3184 &    $0.112 \pm 0.017$ &    0.034 &            0.70 \\
NGC 3351 &    $0.08 \pm 0.02$ &      0 &                0.69 \\
NGC 3938 &    $0.12 \pm 0.02$ &      0 &                0.82 \\
NGC 4254 &    $0.093 \pm 0.016$ &    0.013 &            0.83 \\
NGC 4321 &    $0.113 \pm 0.015$ &    0 &                0.78 \\
NGC 4579 &    $0.11 \pm 0.06$ &      0 &                0.51 \\
NGC 4725 &    $0.07 \pm 0.03$ &      0.17 &             0.12 \\ 
NGC 5055 &    $0.111 \pm 0.011$ &    0.26 &             0.80 \\
NGC 5194 &    $0.159 \pm 0.012$ &    0.56 &             0.80 \\
NGC 6946 &    $0.179 \pm 0.013$ &    0.040 &            0.83 \\
NGC 7793 &    $0.300 \pm 0.019$ &    0 &                0.88 \\
\hline
\end{tabular}
\end{center}
$^a$ These slopes and intrinsic scatter measurements are for the lines
     fit to the data in Figure~\ref{f_pahvs160}.  The data are
     measured within 45~arcsec bins in images with the same resolution
     as the 160~$\mu$m images. Intrinsic scatter measurements of 0 are reported
     when the value calculated with Equation~\ref{e_scatter} is
     negative; this indicates that measurement uncertainties can
     account for all of the scatter in the relation.\\
$^b$ The Spearman's rank correlation coefficient may have values between
     -1 and 1.  A value close to 1 indicates a direct correlation between
     two values.  A value close to -1 indicates an inverse correlation.  A
     value of 0 indicates no correlation.
\end{table}
\renewcommand{\thefootnote}{\arabic{footnote}}

Figure~\ref{f_pahvs160} shows how the (PAH 8~$\mu$m)/160~$\mu$m
surface brightness ratio varies with 160~$\mu$m surface brightness
among the sample galaxies, and the slopes and intrinsic scatter for the
best fit lines as well as Spearman's correlation coefficients for the
data are given in Table~\ref{t_pahvs160}.  Again, the best fitting
lines are determined using uncertainties in both the x- and
y-directions to weight the data.

For all galaxies in the sample, the (PAH 8~$\mu$m)/160~$\mu$m ratio
generally increases as the 160~$\mu$m surface brightness increases, although
the slopes of the relations are relatively shallow for some galaxies,
such as NGC~3031, NGC~3351, and NGC~4725.  If the slopes of the best
fit lines in Figure~\ref{f_pahvs160} were equivalent to 0, this would
indicate that a one-to-one correspondence exists between the PAH 8 and
160~$\mu$m bands.  However, since the slopes are instead all
positive, this indicates that the colours change from low to high
surface brightness regions.

\begin{figure*}
\epsfig{file=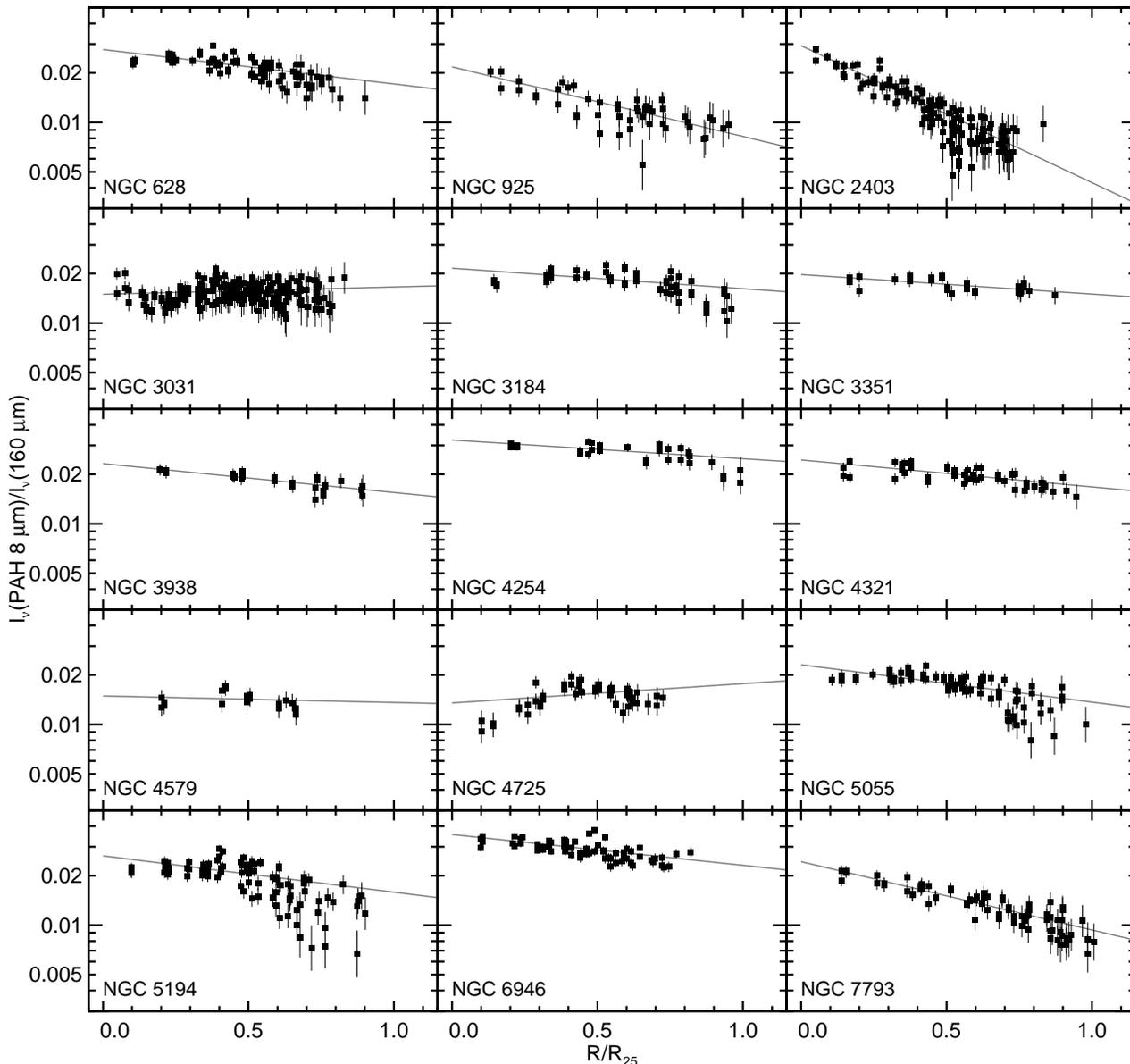}
\caption{Plots of the (PAH 8~$\mu$m)/160~$\mu$m surface brightness
ratios versus deprojected galactocentric radii for the 45~arcsec
square regions measured in these galaxies.  The radii are normalized
by the radius of the D$_{25}$ isophote given by \citet{ddcbpf91}.  The
grey lines are the best fitting lines for the relations in each plot;
slopes and intrinsic scatter for these fits are given in
Table~\ref{t_pah160vsdist}.  Only the uncertainties in the (PAH
8~$\mu$m)/160~$\mu$m ratio are used to weight the data in the fit; the
uncertainties in the radius are assumed to be negligible in this
analysis.}
\label{f_pah160vsdist}
\end{figure*}

The scatter in the data around the best fit lines generally appears to
be at the 10\%-20\% level in many cases.  According to the intrinsic
scatter measurement used here, the scatter in many of the plots can be
explained mostly by uncertainties in the measurements.  For most
galaxies, the intrinsic scatter measurements in Table~\ref{t_pahvs160}
are either similar to or notably lower than the values in
Table~\ref{t_pahvs24}.  Because the data used for
Tables~\ref{t_pahvs24} and \ref{t_pahvs160} were measured in images
that were degraded to the resolution of the 160~$\mu$m images,
resolution effects should not be a factor in this comparison.  Hence,
this comparison between the intrinsic scatter measurements
demonstrates quantitatively that the relation between PAH 8 and
160~$\mu$m emission may exhibit less scatter than the relation between
PAH 8 and 24~$\mu$m emission.

Also note that very low and very high surface brightness 45~arcsec
regions in NGC~5194 and NGC~5055 fall below the best fit line. A
related phenomenon is visible in NGC~4725, where the 45~arcsec regions
within the inner ring fall below the best fit line in
Figure~\ref{f_pahvs160}.  The disparity in the slopes between the high
and low surface brightness data for some galaxies demonstrates that
the (PAH 8~$\mu$m)/160~$\mu$m ratio either stops rising or decreases
in the high surface brightness centres of the galaxies, as can also be
seen in the maps of the (PAH 8~$\mu$m)/160~$\mu$m ratio in
Figure~\ref{f_map} and in the plots of the (PAH 8~$\mu$m)/160~$\mu$m
ratio versus radius in Figure~\ref{f_pah160vsdist}.

\begin{table}
\begin{center}
\caption{Slopes and Intrinsic Scatter for the Best Fit Line Describing (PAH
8)~/~160~$\mu$m Surface Brightness Ratio versus Radius$^a$
\label{t_pah160vsdist}}
\begin{tabular}{@{}lccccc@{}}
\hline
Name &        Slope &                Intrinsic &   Spearman's Rank \\
&             &                      Scatter &     Correlation \\
&             &                      &             Coefficient$^b$ \\
\hline
NGC 628 &     $-0.21 \pm 0.03$ &     0.036 &       -0.80 \\
NGC 925 &     $-0.42 \pm 0.04$ &     0.038 &       -0.74 \\
NGC 2403 &    $-0.83 \pm 0.03$ &     0.29 &        -0.88 \\
NGC 3031 &    $0.05 \pm 0.02$ &      0 &           0.11 \\
NGC 3184 &    $-0.125 \pm 0.03$ &    0.11 &        -0.68 \\
NGC 3351 &    $-0.12 \pm 0.04$ &     0 &           -0.68 \\
NGC 3938 &    $-0.18 \pm 0.04$ &     0 &           -0.84 \\
NGC 4254 &    $-0.11 \pm 0.02$ &     0.039 &       -0.74 \\
NGC 4321 &    $-0.16 \pm 0.02$ &     0.00177 &     -0.73 \\
NGC 4579 &    $-0.04 \pm 0.08$ &     0 &           -0.32 \\
NGC 4725 &    $0.12 \pm 0.05$ &      0.130 &       0.18 \\
NGC 5055 &    $-0.23 \pm 0.02$ &     0.28 &        -0.79 \\
NGC 5194 &    $-0.22 \pm 0.02$ &     1.07 &        0.71 \\
NGC 6946 &    $-0.188  \pm 0.016$ &  0.069 &       -0.73 \\
NGC 7793 &    $-0.42 \pm 0.03$ &     0 &           -0.90 \\
\hline
\end{tabular}
\end{center}
$^a$ These slopes and intrinsic scatter measurements are for the lines
     fit to the data in Figure~\ref{f_pah160vsdist}.  The data are
     measured within 45~arcsec bins in images with the same resolution
     as the 160~$\mu$m images.  Intrinsic scatter measurements of 0
     are reported when the value calculated with
     Equation~\ref{e_scatter} is negative; this indicates that
     measurement uncertainties can account for all of the scatter in
     the relation.\\
$^b$ The Spearman's rank correlation coefficient may have values between
     -1 and 1.  A value close to 1 indicates a direct correlation between
     two values.  A value close to -1 indicates an inverse correlation.  A
     value of 0 indicates no correlation.
\end{table}

Figure~\ref{f_map} illustrates how the (PAH 8~$\mu$m)/160~$\mu$m ratio
may peak outside the nuclei of nearby galaxies.  From the ratio maps
alone, it is apparent that the (PAH 8~$\mu$m)/160~$\mu$m ratio does
not necessarily monotonically decrease from the nuclei to the edges of
the optical discs as was suggested by \citet{betal06}.  Both
Figure~\ref{f_pah160vsdist}, which plots the (PAH 8~$\mu$m)/160~$\mu$m
ratio versus deprojected galactocentric radius for 45~arcsec regions
in these galaxies, and Table~\ref{t_pah160vsdist}, which gives the
slopes and intrinsic scatter measurements for the best fit lines in
Figure~\ref{f_pah160vsdist} as well as the Spearman's correlation
coefficient for the relations, support this conclusion.  First, note
that the regions with the highest (PAH 8~$\mu$m)/160~$\mu$m ratios
within some of these galaxies are found outside the nucleus.  As can
be seen in Figure~\ref{f_map}, the regions with enhanced (PAH
8~$\mu$m)/160~$\mu$m ratios may correspond to spiral structure, as is
most clearly seen in NGC~3031 and NGC~6946.  In NGC~4725, the inner
ring has the highest (PAH 8~$\mu$m)/160~$\mu$m ratio, not the nucleus.
The intrinsic scatter in the fits versus surface brightness is
generally lower than for those for the fits versus radius.  Moreover,
the Spearman's correlation coefficients for the relations between the
(PAH 8~$\mu$m)/160~$\mu$m ratio and 160~$\mu$m surface brightness
generally has a higher absolute value than the corresponding
correlation coefficients for the relation between the (PAH
8~$\mu$m)/160~$\mu$m ratio and radius.  These results suggest that the
ratio may be more strongly dependent on 160~$\mu$m surface brightness
than radius.

Since the (PAH 8~$\mu$m)/160~$\mu$m ratio may be dependent on dust
heating, we also examine how the ratio is related to the
24~$\mu$m/160~$\mu$m ratio.  The 24~$\mu$m band, which traces
$\gtrsim100$~K hot dust emission, increases faster than other infrared
band (including the PAH 8 and 160~$\mu$m bands) as the illuminating
radiation field increases \citep{dhcsk01, ld01, dl07}.  The 160~$\mu$m
band is approximately directly proportional to the TIR flux, as
discussed in Section~\ref{s_data_band}.  Therefore, the
24~$\mu$m/160~$\mu$m ratio should be a reasonable indicator of dust
heating.  If the mass fraction of PAHs remains constant and if the
(PAH 8~$\mu$m)/160~$\mu$m ratio is dependent on dust heating, then the
(PAH 8~$\mu$m)/160~$\mu$m ratio should monotonically increase as the
24~$\mu$m/160~$\mu$m ratio increases, although the slope may not
necessarily be constant.  This comparison is similar to the direct
comparison between the PAH 8 and 24~$\mu$m bands performed in
Section~\ref{s_comp_pah24}, but the normalisation with the 160~$\mu$m
band removes variations related to dust surface density.

\begin{figure*}
\epsfig{file=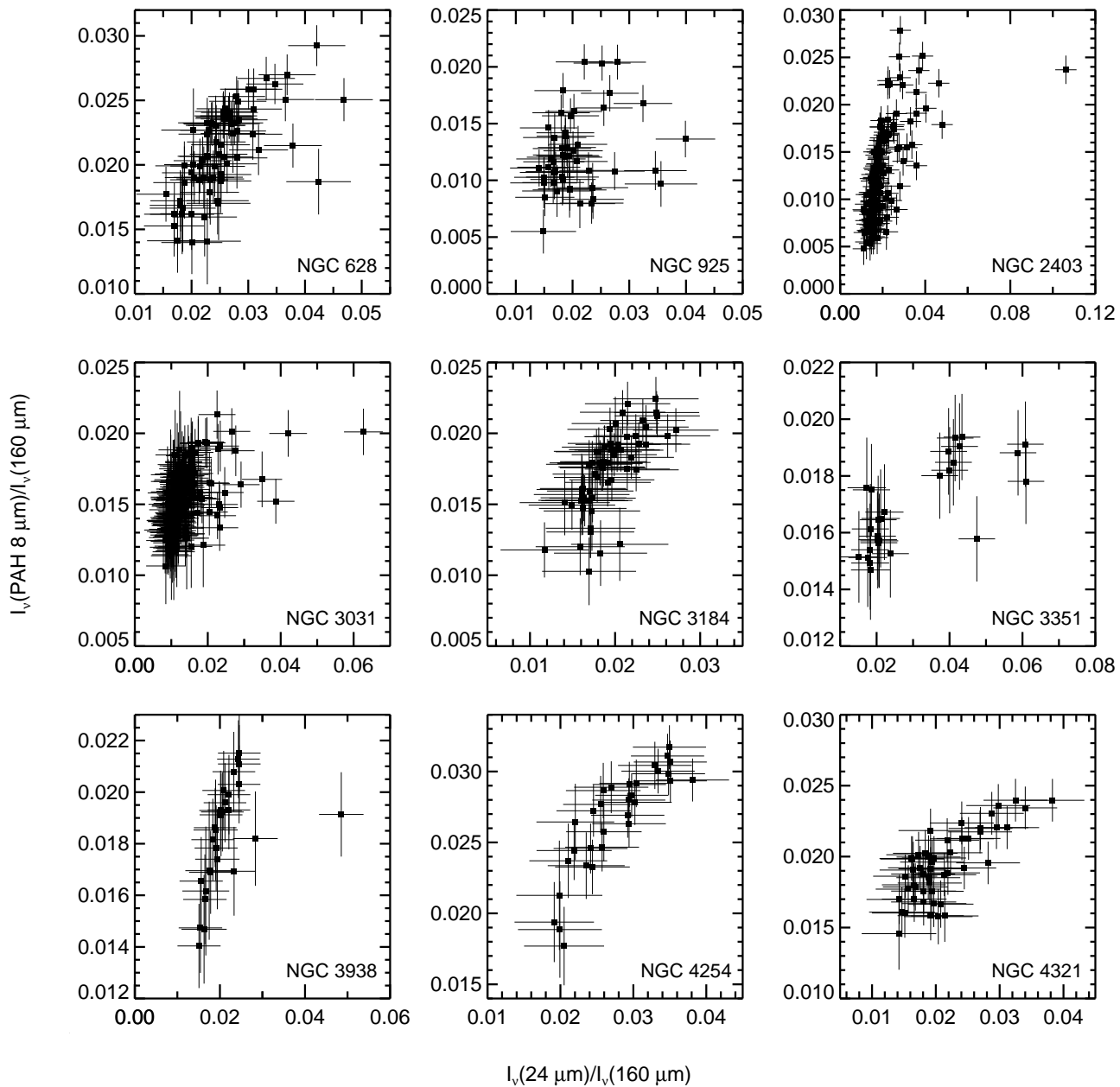}
\caption{Plots of the (PAH 8~$\mu$m)/160~$\mu$m surface brightness ratios
versus the 24~$\mu$m/160~$\mu$m surface brightness ratios for the 45~arcsec
square regions measured in these galaxies.  The 24~$\mu$m/160~$\mu$m ratios
are used as a proxy for dust heating here.}
\label{f_pah160vs24160}
\end{figure*}

\begin{figure*}
\epsfig{file=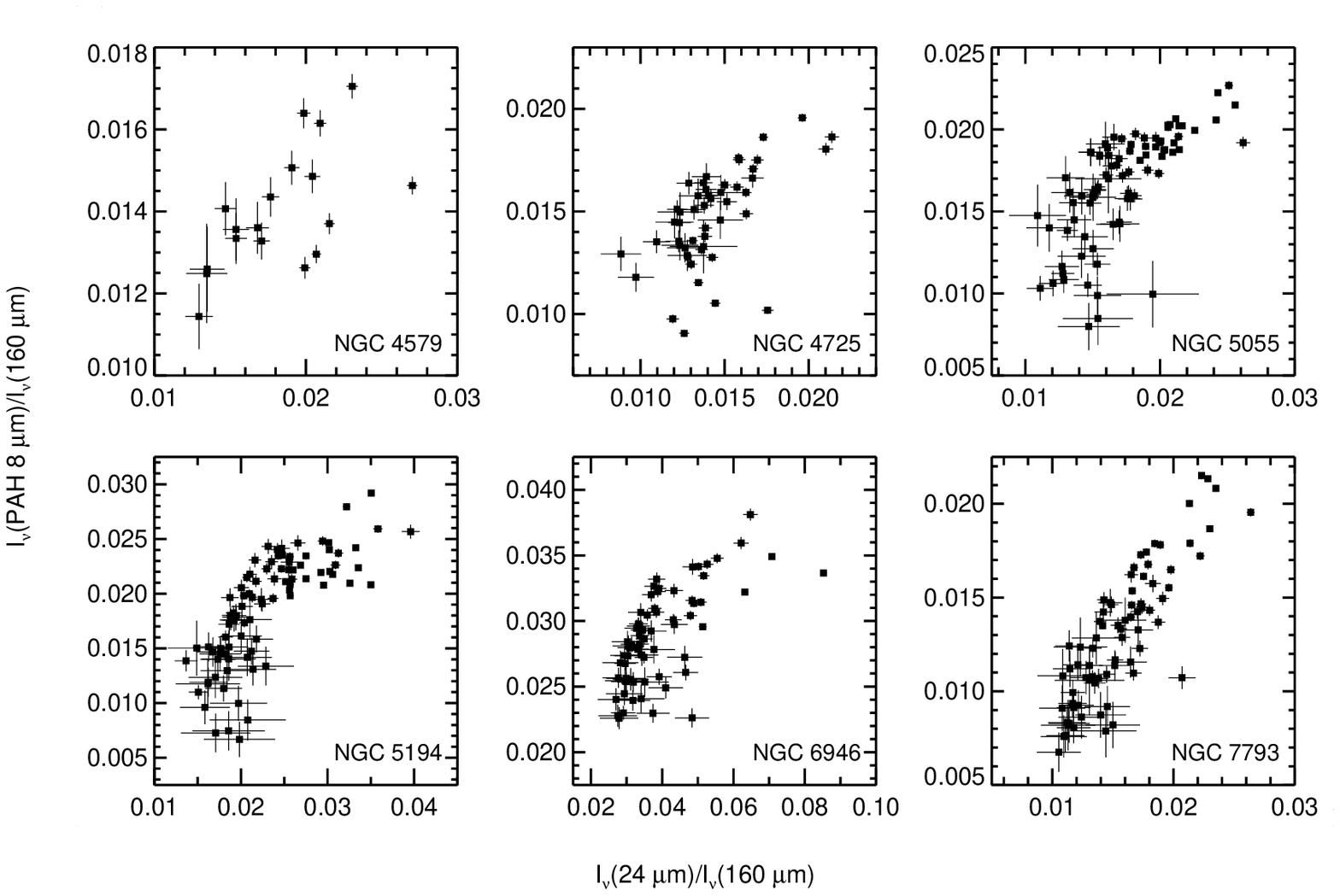}
\contcaption{}
\end{figure*}

\begin{figure*}
\epsfig{file=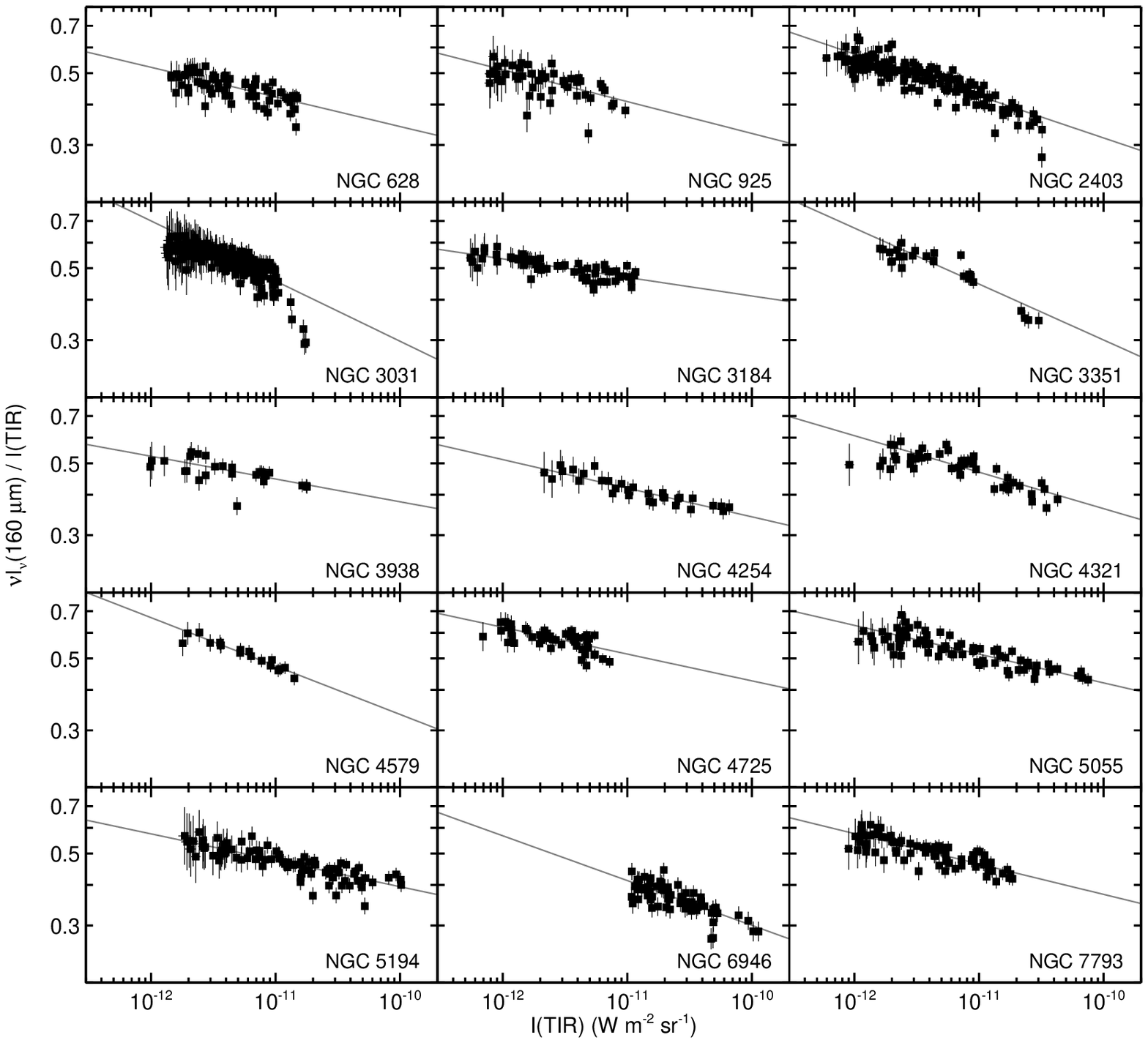}
\caption{Plots of the 160~$\mu$m/TIR surface brightness ratios
versus TIR surface brightness for the 45~arcsec square regions
measured in these galaxies.  The grey lines are the best fitting lines
for the relations in each plot; slopes for these fits are given in
Table~\ref{t_160vstir}.  Note that the uncertainties in the x- and
y-directions are used to weight the data in the fit.}
\label{f_160vstir}
\end{figure*}

\begin{figure*}
\epsfig{file=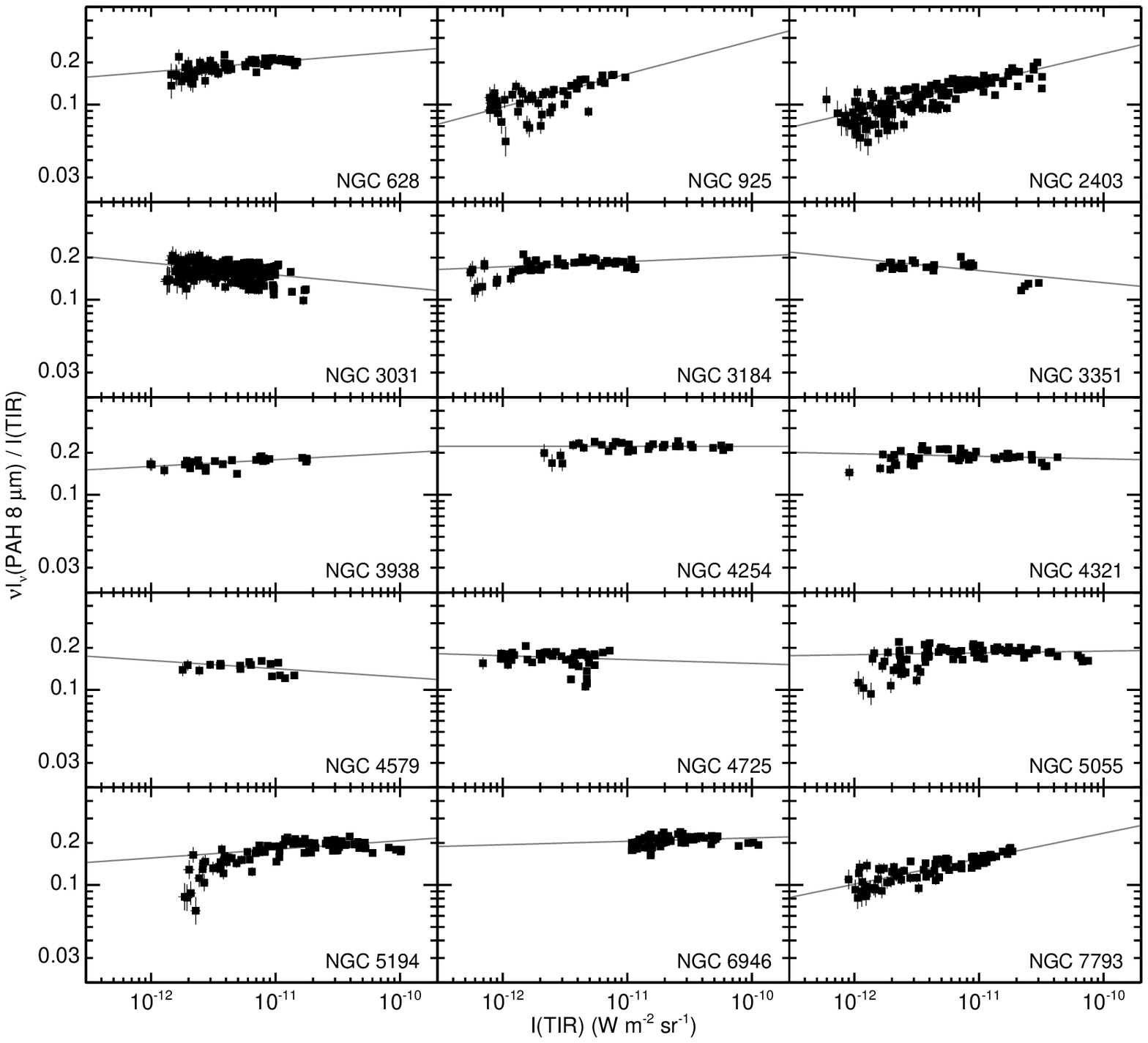}
\caption{Plots of the (PAH 8~$\mu$m)/TIR surface brightness ratios
versus TIR surface brightness for the 45~arcsec square regions
measured in these galaxies.  The grey lines are the best fitting lines
for the relations in each plot; slopes for these fits are given in
Table~\ref{t_pahvstir}.  Note that the uncertainties in the x- and
y-directions are used to weight the data in the fit.}
\label{f_pahvstir}
\end{figure*}

Figure~\ref{f_pah160vs24160} shows how the (PAH 8~$\mu$m)/160~$\mu$m
ratio varies with the 24~$\mu$m/160~$\mu$m ratio within 45~arcsec
regions in the sample galaxies.  Many of the 45~arcsec regions that
are relatively weak in 24~$\mu$m emission tend to show a tight
correspondence between the (PAH 8~$\mu$m)/160~$\mu$m and
24~$\mu$m/160~$\mu$m ratios, but some regions with enhanced 24~$\mu$m
emission appear far to the right of these curves.  The most
spectacular example is NGC~3938, where the ratios for most of the
regions closely follow a linear relation but the H{\small II} region
on the east side of the disc falls far to the right of this relation.
The relation between the two ratios on the left side of the panels in
Figure~\ref{f_pah160vs24160} suggest that the (PAH
8~$\mu$m)/160~$\mu$m ratio is dependent on dust heating, but the
outliers on the right in these panels show PAH 8~$\mu$m emission is
not enhanced in areas with strong dust heating such as H{\small II}
regions.  Further interpretation of these results is presented in the
next section.

For another perspective on the nature of the variation in the (PAH
8~$\mu$m)/160~$\mu$m ratio, we examined how the 160~$\mu$m/TIR and
(PAH 8~$\mu$m)/TIR ratios vary as a function of TIR surface
brightness.  These are displayed in Figures~\ref{f_160vstir} and
\ref{f_pahvstir}, with slopes for the best fit lines given in
Table~\ref{t_160vstir} and Table~\ref{t_pahvstir}.  As stated in
Section~\ref{s_data_band}, the 160~$\mu$m/TIR ratio should gradually
decrease as dust temperatures increase above $\sim15$~K.  This is
reflected in the general trends visible in Figure~\ref{f_160vstir}.
Some scatter is visible in this figure, but this is expected, as the
ratio will deviate from this general trend where 24~$\mu$m emission is
strongly enhanced within star forming regions.  The (PAH 8~$\mu$m)/TIR
ratio is constant or changes very little in many galaxies (e.g. the
slope of the best fit line in Figure~\ref{f_pahvstir} deviates less
than $3\sigma$ from 0), which indicates that the (PAH~8$\mu$m)
emission is directly proportional to TIR emission in these galaxies.
Some scatter related to the scatter seen in Figure~\ref{f_pahvs24} is
visible in these data, and in some cases, the infrared-brightest
regions near the nucleus deviate from the trend in (PAH 8~$\mu$m)/TIR
ratio versus TIR surface brightness, which makes the best fit lines
appear positive.  In a few exceptional galaxies, however, the (PAH
8~$\mu$m)/TIR ratio increases as TIR surface brightness increases.  We
conclude that the variations in (PAH 8~$\mu$m)/160~$\mu$m ratio are
driven by changes in 160~$\mu$m emission relative to total dust
emission in all galaxies and also by changes in (PAH 8~$\mu$m)
emission relative to total dust emission in some situations.  This is
discussed further in the following section.

\section{Discussion}
\label{s_discuss}

\subsection{Interpretation of the relation between PAH 8 and 24~$\mu$m 
emission}
\label{s_discuss_pah24}

The results in Section~\ref{s_comp_pah24} as well as
\ref{s_comp_pah160} demonstrate that the relation between PAH 8 and
24~$\mu$m emission exhibits a significant scatter on spatial scales
smaller than $\sim2$~kpc.  In contrast, some ISO results from
comparisons of 7 and 15~$\mu$m emission had implied that the ratio of
PAH to hot dust emission should be relatively uniform across the discs
of most galaxies \citep[e.g.][]{retal01}.  However, some ISO studies
actually found significant variations in the 7~$\mu$m/15~$\mu$m ratio
\citep{hkb02}, and even studies that did find uniform
7~$\mu$m/15~$\mu$m colours within the discs of galaxies noted that the
ratio might decrease within starbursts in the centres of some galaxies
\citep{retal01}.  The variations in the (PAH 8~$\mu$m)/24~$\mu$m ratio
observed in this subsample of SINGS galaxies are consistent with {\it
Spitzer} results for individual galaxies, such as for M51
\citep{cetal05}, M81 \citep{petal06}, and NGC~4631 \citep{betal06}.
Moreover, we find that 24~$\mu$m emission is more peaked in the
centres of H{\small II} regions while the PAH 8~$\mu$m emission is
relatively stronger outside H{\small II} regions, which is consistent
with similar phenomena observed in NGC~300 \citep{hraetal04},
NGC~4631 \citep{betal06}, and M101 \citep{getal08}.

The variations in the (PAH 8~$\mu$m)/24~$\mu$m ratio are best seen by
comparing bright star-forming regions and diffuse regions.  If,
because of the limited sensitivity of the data, the ratio between PAH
and hot dust emission is only measured in the bright regions in some
galaxies, as was done by \citet{retal01}, then the ratio between PAH
and hot dust emission may appear uniform.  When these bright regions
are compared to diffuse emission, however, variations may be seen in
the ratio of PAH to hot dust emission.  This may be the primary reason
why some results from ISO suggested that the ratio of PAH to hot dust
emission was uniform in most spiral galaxies whereas variations may be
seen in {\it Spitzer} data.

\begin{table}
\begin{center}
\renewcommand{\thefootnote}{\alph{footnote}}
\caption{Results for the Best Fit Line Describing 160~$\mu$m~/~TIR
  Surface Brightness Ratio versus TIR Surface Brightness
  \label{t_160vstir}}
\begin{tabular}{@{}lccc@{}}
\hline
Name &        Slope$^a$ \\
\hline
NGC 628 &     $-0.092 \pm 0.011$ \\
NGC 925 &     $-0.098 \pm 0.017$ \\
NGC 2403 &    $-0.130 \pm 0.005$ \\
NGC 3031 &    $-0.186 \pm 0.012$ \\
NGC 3184 &    $-0.057 \pm 0.009$ \\
NGC 3351 &    $-0.173 \pm 0.013$ \\
NGC 3938 &    $-0.070 \pm 0.015$ \\
NGC 4254 &    $-0.088 \pm 0.014$ \\
NGC 4321 &    $-0.113 \pm 0.009$ \\
NGC 4579 &    $-0.15 \pm 0.02$ \\
NGC 4725 &    $-0.083 \pm 0.013$ \\
NGC 5055 &    $-0.089 \pm 0.006$ \\
NGC 5194 &    $-0.082 \pm 0.008$ \\
NGC 6946 &    $-0.138 \pm 0.013$ \\
NGC 7793 &    $-0.094 \pm 0.009$ \\
\hline
\end{tabular}
\end{center}
$^a$ These slopes are for the lines fit to the data in
     Figure~\ref{f_160vstir}.  The data are measured within 45~arcsec
     bins in images with the same resolution as the 160~$\mu$m
     images.\\
\end{table}
\renewcommand{\thefootnote}{\arabic{footnote}}

\begin{table}
\begin{center}
\renewcommand{\thefootnote}{\alph{footnote}}
\caption{Results for the Best Fit Line Describing (PAH
  8~$\mu$m)~/~TIR Surface Brightness Ratio versus TIR Surface
  Brightness \label{t_pahvstir}}
\begin{tabular}{@{}lccc@{}}
\hline
Name &        Slope$^a$ \\
\hline
NGC 628 &     $0.073 \pm 0.012$ \\
NGC 925 &     $0.237 \pm 0.018$ \\
NGC 2403 &    $0.210 \pm 0.006$ \\
NGC 3031 &    $-0.086 \pm 0.011$ \\
NGC 3184 &    $0.037 \pm 0.009$ \\
NGC 3351 &    $-0.087 \pm 0.012$ \\
NGC 3938 &    $0.048 \pm 0.012$ \\
NGC 4254 &    $0.000 \pm 0.010$ \\
NGC 4321 &    $-0.018 \pm 0.007$ \\
NGC 4579 &    $-0.06 \pm 0.03$ \\
NGC 4725 &    $-0.028 \pm 0.013$ \\
NGC 5055 &    $0.013 \pm 0.006$ \\
NGC 5194 &    $0.062 \pm 0.006$ \\
NGC 6946 &    $0.024 \pm 0.008$ \\
NGC 7793 &    $0.183 \pm 0.010$ \\
\hline
\end{tabular}
\end{center}
$^a$ These slopes are for the lines fit to the data in
     Figure~\ref{f_pahvstir}.  The data are measured within 45~arcsec
     bins in images with the same resolution as the 160~$\mu$m
     images.\\
\end{table}
\renewcommand{\thefootnote}{\arabic{footnote}}

Additionally, differences between the nature of 15 and 24~$\mu$m dust
emission could have also contributed to the differences between the
7~$\mu$m/15~$\mu$m relation observed with ISO and the (PAH
8~$\mu$m)/24~$\mu$m relation observed with {\it Spitzer}.  Based on
semi-empirical and theoretical models, the 24~$\mu$m band is expected
to increase faster than the 15~$\mu$m as the illuminating radiation
field increases \citep{dhcsk01, ld01, dl07}.  Moreover, unlike the
24~$\mu$m band, a significant fraction of the 15~$\mu$m band includes
PAH emission \citep{setal07, dl07}.  Consequently, a comparison of PAH
to 24~$\mu$m data should exhibit more scatter than a comparison of PAH
to 15~$\mu$m data.

We also found that differences in the distribution of star-forming
regions within galaxies could influence how the (PAH
8~$\mu$m)/24~$\mu$m ratio varies with 24~$\mu$m surface brightness.
In some galaxies, such as NGC~3351 and NGC~6946, the nucleus is the
strongest site of star formation.  Hence, the location expected to
have the lowest (PAH 8~$\mu$m)/24~$\mu$m ratio will correspond to the
location with the highest 24~$\mu$m surface brightness, so the data
will show that the (PAH 8~$\mu$m)/24~$\mu$m ratio decreases as the
24~$\mu$m surface brightness increases, thus varying as expected from
many models of dust emission \citep[e.g.][]{dhcsk01, ld01, dl07}.  In
other galaxies, such as NGC~925 and NGC~2403, the nucleus is not the
site of the strongest star formation activity, and many H{\small II}
regions can be found at the periphery of the regions that were
detected at the $3\sigma$ level in these data.  While point-like
24~$\mu$m sources should correspond to H{\small II} regions
\citep{pkbetal07, cetal07}, dust emission models such as those
presented by \citet{ld01}, \citet{dhcsk01}, and \citet{dl07} suggest
that diffuse 24~$\mu$m dust emission may still potentially originate
from regions outside star-forming regions with high radiation fields.
Furthermore, ionising photons may escape from star-forming regions
into the diffuse interstellar medium \citep{oetal07}, and results from
\citet{cetal07} suggest that 24~$\mu$m emission may trace the
recombination line emission from the diffuse gas heated by these
photons.  Hence, it is possible that the infrared-brightest regions in
the centres of some galaxies may contain a higher fraction of
24~$\mu$m emission from diffuse dust than the infrared-faint regions,
so the (PAH 8~$\mu$m)/24~$\mu$m ratio will increase as the 24~$\mu$m
surface brightness increases, as can be seen for NGC~925 and NGC~2403
in Figure~\ref{f_pahvs24}.

Several mechanisms may be responsible for creating the differences in
the spatial distribution of the PAH 8 and 24~$\mu$m emission.  First,
as the illuminating radiation field increases, the 24~$\mu$m band is
thought to increase more rapidly than bands that trace PAH emission
\citep{dhcsk01, ld01, dl07}.  Therefore, the 24~$\mu$m emission may be
more strongly enhanced than the PAH emission in regions with very high
radiation fields, such as star-forming regions.  Second, the PAHs may
be destroyed in regions with strong radiation fields or in regions
with large numbers of high-energy photons \citep[e.g.][]{metal06}.
The PAHs would be absent, and the (PAH 8~$\mu$m)/24~$\mu$m ratio would
be low in the centres of star-forming regions and AGN.  Third, the
ratio of the 7.7~$\mu$m PAH feature to other PAH features may vary
with changes in the charge state of the PAHs \citep[e.g.][]{ahs99,
ld01, dl07}.  This could occur in H{\small II} regions if the electron
densities are high enough that recombination rates becomes significant
compared to photoionisation rates, which would lead to a reduction of
the PAH$^+$ ions that produce the 7.7~$\mu$m emission \citep{wd01}.

While the enhancement of 24~$\mu$m emission linked to dust heating in
star-forming regions is certainly at least part of the reason why the
(PAH 8~$\mu$m)/24~$\mu$m ratio varies, we argue that the variations
must also be caused in part by a reduction in PAH emission in the
8~$\mu$m band through either PAH destruction or changes in the
relative strengths of PAH spectral features.  Models of dust emission
have suggested that, if PAHs are present and radiating at
8~$\mu$m, the (PAH 8~$\mu$m)/160~$\mu$m ratio would be enhanced in
regions with very strong radiation fields, warmer dust, and low (PAH
8~$\mu$m)/24~$\mu$m ratios \citep{dl07}.  However, the (PAH
8~$\mu$m)/160~$\mu$m ratio does not peak within individual
star-forming regions where local minima in the (PAH
8~$\mu$m)/24~$\mu$m ratio are found, such as the bright extranuclear
regions in NGC~2403, NGC~3184, and NGC~3938.  This is best
demonstrated by comparisons of the maps of (PAH 8~$\mu$m)/24~$\mu$m
and (PAH 8~$\mu$m)/160~$\mu$m ratios at matching resolutions in
Figure~\ref{f_map_conv160comp}, although the additional discussion in
Section~\ref{s_comp_pah160} and the maps in Figure~\ref{f_map} also
support this conclusion.  Furthermore, Figure~\ref{f_pah160vs24160}
demonstrates that the (PAH 8~$\mu$m)/160~$\mu$m ratios are not always
higher in regions with strongly enhanced 24~$\mu$m emission.
According to the models of \citet{dl07}, the (PAH 8~$\mu$m)/160~$\mu$m
should monotonically increase with the 24~$\mu$m/160~$\mu$m ratio in
these regions if PAHs are present and if the PAH ionisation does not
change.  These results indicate that the variations in the (PAH
8~$\mu$m)/24~$\mu$m ratio must be in part caused by the suppression of
PAH 8~$\mu$m emission.

Observations of individual H{\small II} regions within the Milky Way
have shown that the PAH emission may be found primarily in shell-like
structures around the star-forming regions \citep[e.g.][]{rrlf06,
cetal06, sb07} and that the strength of all PAH spectral features
relative to 24~$\mu$m hot dust emission decreases within the centres
of H{\small II} regions \citep[e.g.][]{pscetal07,lbbdh07}.  If these
results are applicable to the galaxies in this paper, then the
inferred decrease in the (PAH 8~$\mu$m)/24~$\mu$m ratio within
star-forming regions may also be partly caused by PAH destruction and
not just variations in the strengths of PAH spectral features.
Further work with mid-infrared spectroscopic observations of
star-forming regions within these galaxies, specifically studies of
how the total emission in all PAH spectral feature emission varies
with respect to total dust emission, would be needed to confirm that
PAH destruction is taking place.

Although the PAH 8~$\mu$m emission does not share a one-to-one
correspondence with 24~$\mu$m emission or with other star formation
tracers on kiloparsec scales, this does not necessarily preclude its
use as a tracer of integrated star formation within nearby spiral
galaxies.  The variations between PAH 8~$\mu$m emission and other star
formation tracers could potentially be averaged out when integrating
across the optical discs of spiral galaxies.  However, because a
significant fraction of the PAH 8~$\mu$m emission still originates
from the diffuse ISM, it would not be as reliable as other global star
formation tracers.  Furthermore, because the ratio of PAH to TIR flux
and the ratio of PAH to 24~$\mu$m flux density varies with metallicity
\citep{eetal05, detal05, ddbetal07, cetal07, eetal08}, a conversion
factor relating integrated PAH 8~$\mu$m emission to star formation may
also vary with global metallicity.

\begin{figure}
\epsfig{file=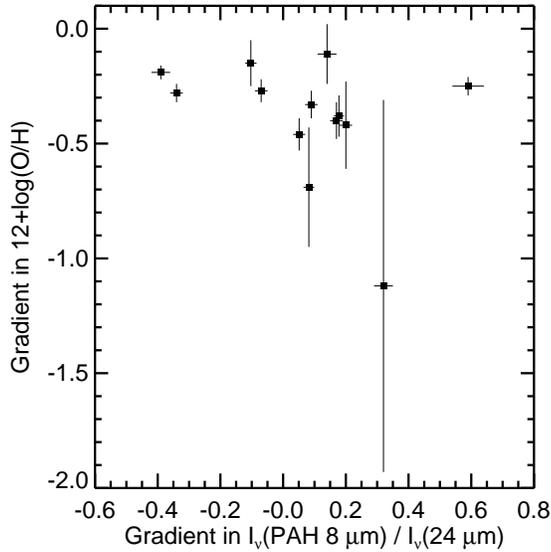}
\caption{Plot of the radial gradients in the (PAH 8~$\mu$m)/24~$\mu$m
ratio from Table~\ref{t_pah24vsdist} versus radial gradients in
12+log(O/H) measured by J. Moustakas et al. (2008, in preparation)
using the \citet{pt05} calibration.  The gradients are in units of dex
divided by the radius of the D$_{25}$ isophote.  Gradients in
12+log(O/H) were not provided for NGC~3938 and NGC~4579.}
\label{f_gradcomp_pah24_12logoh}
\end{figure}

We also note that variations in metallicity could possibly produce
variations in the (PAH 8~$\mu$m)/24~$\mu$m ratio observed in the
galaxies studied in this paper, but these variations would not be
expected to cause the differences in the ratio between star-forming
regions and the diffuse ISM observed here.  Moreover, the gas-phase
abundances of the galaxies in our sample are well above
12+log(O/H)$\simeq$8-8.1, where PAH 8~$\mu$m emission should be
suppressed as determined by \citet{ddbetal07} and \citet{eetal08},
although NGC~628, NGC~925, NGC~2403, and NGC~3031 all have
characteristic abundances close to this value and might be affected.
To examine this issue further, we compared the radial gradients in the
(PAH 8~$\mu$m)/24~$\mu$m ratio to the radial gradients in 12+log(O/H)
measured by J. Moustakas et al. (2008, in preparation) using the
\citet{pt05} calibration for 13 of the galaxies in our sample.
(Abundance gradients are not given by J. Moustakas et al. for NGC~3938
and NGC~4579, and so these galaxies are not included in this
comparison.)  The relation between these two radial gradients can be
seen in Figure~\ref{f_gradcomp_pah24_12logoh}.  If the variations in
the (PAH 8~$\mu$m)/24~$\mu$m ratio depended primarily on metallicity
variations, then we would expect to see a correlation between the
gradients in Figure~\ref{f_gradcomp_pah24_12logoh}.  However, the two
gradients appear uncorrelated, which indicates that the variations in
the (PAH 8~$\mu$m)/24~$\mu$m ratio that have been observed here depend
primarily on effects unrelated to metallicity.  This conclusion for
these spiral galaxies is also consistent with the results presented
for the spiral galaxy M101 by \citet{getal08}, who found that PAH
equivalent widths appeared more dependent on the ionisation of the ISM
(characterised using [Ne{\small III}]/[Ne{\small II}] and [S{\small
IV}]/[S{\small III}]) than 12+log(O/H).

\subsection{Interpretation of the relation between PAH 8 and 160~$\mu$m 
emission}

The (PAH 8~$\mu$m)/160~$\mu$m ratio appears to be closely correlated
with the 160~$\mu$m surface brightness measured on 45~arcsec scales in
most galaxies, even in nearby galaxies where this angular scale
corresponds to $\ltsim 1$~kpc.  Moreover, the (PAH
8~$\mu$m)/160~$\mu$m ratio sometimes traces high surface brightness
large scale structures in the discs of these galaxies, such as the
spiral arms in NGC~3031 or NGC~6946.  This indicates that the
variations in the ratio may not be primarily dependent on radius as
inferred by \citet{betal06} but instead may be primarily dependent
on the 160~$\mu$m surface brightness.  However, the large scale
structures within these galaxies are only marginally resolved in the
160~$\mu$m data.  Higher resolution observations at wavelengths longer
than 100~$\mu$m are needed to confirm that this interpretation is
valid.

\begin{figure}
\epsfig{file=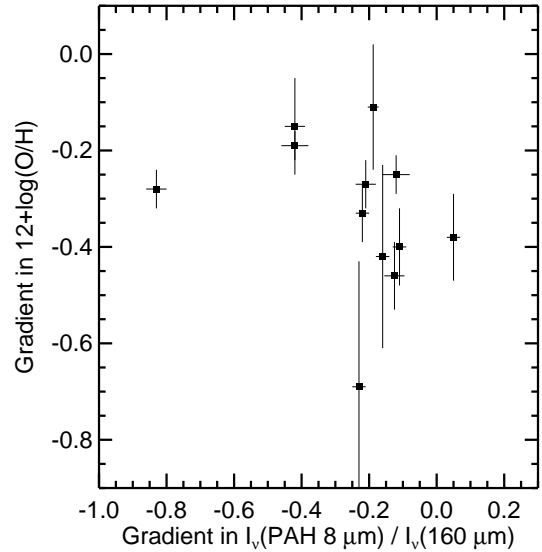}
\caption{Plot of the radial gradients in the (PAH 8~$\mu$m)/160~$\mu$m
ratio from Table~\ref{t_pah160vsdist} versus radial gradients in
12+log(O/H) measured by J. Moustakas et al. (2008, in preparation)
using the \citet{pt05} calibration.  The gradients are in units of dex
divided by the radius of the D$_{25}$ isophote.  Gradients in
12+log(O/H) were not provided for NGC~3938 and NGC~4579, and the
gradients in the (PAH 8~$\mu$m)/160~$\mu$m ratio for NGC~4725 changes
significantly between the centre and outer disc, so these three
galaxies are not included in this plot.}
\label{f_gradcomp_pah160_12logoh}
\end{figure}

Nonetheless, if the above interpretation is correct, then it also
suggests that the variations in the (PAH 8~$\mu$m)/160~$\mu$m ratio
may be more dependent on 160~$\mu$m surface brightness than
metallicity within the regions studied in these galaxies.  Since
variations in metallicity have been linked to decreased PAH emission
relative to longer-wavelength dust emission in the integrated spectra
of galaxies \citep[e.g.][]{eetal05, detal05, ddbetal07, eetal08}, it
was not unreasonable to expect that the observed variations in the
(PAH 8~$\mu$m)/160~$\mu$m ratio within these galaxies might be linked
to metallicity variations.  However, the 12+log(O/H) values measured
in these galaxies by J. Moustakas et al. (2008, in preparation)
generally do not drop below $\sim8$-8.1, which is where
\citet{ddbetal07} and \citet{eetal08} showed that metallicity strongly
affects PAH 8~$\mu$m emission.  Moreover, metallicity is expected to
decrease monotonically with radius, while the (PAH
8~$\mu$m)/160~$\mu$m ratio and the 160~$\mu$m surface brightness do
not, and the metallicity should not peak within substructures such as
spiral arms, whereas the (PAH 8~$\mu$m)/160~$\mu$m ratio and the
160~$\mu$m surface brightness both peak within such substructures.  As
an additional test, we compared the gradients in the (PAH
8~$\mu$m)/160~$\mu$m ratio versus radius with the metallicity
gradients from J. Moustakas et al. (2008, in preparation) calculated
with the \citet{pt05} calibration.  For this comparison, we excluded
NGC~4725 because the gradients in the (PAH 8~$\mu$m)/160~$\mu$m ratio
changes significantly between the nucleus and the outer disc, and we
excluded NGC~3938 and NGC~4579 because abundance gradients are not
given by J. Moustakas et al.  The gradients for the other galaxies are
plotted in Figure~\ref{f_gradcomp_pah160_12logoh}.  If the (PAH
8~$\mu$m)/160~$\mu$m ratio was affected by metallicity in the regions
studied in these galaxies, then these data would be positively
correlated.  Since the data in Figure~\ref{f_gradcomp_pah160_12logoh}
do not exhibit such a correlation, the two gradients may be unrelated.
We therefore conclude that, in the regions of the galaxies studied
here, metallicity variations are not as important as 160~$\mu$m
surface brightness variations in determining the (PAH
8~$\mu$m)/160~$\mu$m ratio, although metallicity may be a factor
outside the optical discs.  Again, this is consistent with the
conclusions reached by \citet{getal08}, who found that PAH equivalent
widths in M101 were more dependent on the ionisation of the ISM than
abundances.

While the (PAH 8~$\mu$m)/160~$\mu$m ratio traces large scale
structure, we have demonstrated that the ratio is not enhanced within
individual star-forming regions, and we explained earlier in this
section that PAH 8~$\mu$m emission must be inhibited in regions with
strong radiation fields.  Based on these conclusions and the strong
relation between the (PAH 8~$\mu$m)/160~$\mu$m ratio and 160~$\mu$m
surface brightness, we infer that the PAHs in these galaxies are
generally associated with the cold ($\sim20$~K) dust that dominates the
160~$\mu$m emission, at least on scales of $\sim2$~kpc.  Because most
of this cold dust may be expected to be found in the diffuse ISM, the
PAHs may also be found primarily in the diffuse ISM as well, although
some of the cold dust and PAHs may also be found within clouds
associated with star-forming regions.  Moreover, since the (PAH
8~$\mu$m)/160~$\mu$m ratio increases as the 160~$\mu$m surface
brightness increases, the (PAH 8~$\mu$m)/160~$\mu$m ratio may be an
indicator of variations in the intensity of the radiation field
heating the diffuse ISM.  This is further supported by the tight
correlation between the (PAH 8~$\mu$m)/160~$\mu$m ratio and the
24~$\mu$m/160~$\mu$m ratio found for many regions with weak 24~$\mu$m
emission in Figure~\ref{f_pah160vs24160}, which are presumably regions
that primarily sample dust emission from the diffuse ISM.
Nonetheless, this interpretation is only valid if the PAH mass
fraction does not vary appreciably between infrared-faint and
infrared-bright regions in the diffuse ISM, and far-infrared
observations with higher angular resolution will be needed to
determine whether this association is still applicable on smaller
spatial scales.

In most galaxies in this sample, the variation in the (PAH
8~$\mu$m)/160~$\mu$m ratio with 160~$\mu$m surface brightness is
purely driven by the decrease in 160~$\mu$m emission relative to TIR
emission, as the (PAH 8~$\mu$m)/TIR ratio remains constant with TIR
surface brightness.  In a few exceptions, however, a decline in the
PAH 8~$\mu$m emission relative to TIR emission is also partly
responsible for the observed variations.  This reduction of PAH
emission may occur if the fraction of starlight from evolved red stars
increases as the 160~$\mu$m surface brightness decreases.  In this
scenario, the photons in the illuminating radiation field would have
reduced energies, and the peak temperatures attained by PAHs following
the absorpiton of single photons would be reduced.  Consequently, the
PAHs would tend to radiate at longer wavelengths, and the (PAH
8~$\mu$m)/160~$\mu$m ratio would appear to decrease.

We also noted that the (PAH 8~$\mu$m)/160~$\mu$m ratio drops within
the high surface brightness centre of some galaxies within this
sample.  A couple of mechanisms could be responsible for this
phenomenon.  First, AGN may be responsible for inhibiting the PAH
8~$\mu$m emission either by changing the ionisation state of the PAHs
or by destroying the PAHs.  This was also suggested by
\citet{setal07}, who showed that the 7.7~$\mu$m PAH spectral feature
was suppressed within the centres of SINGS galaxies with low
luminosity AGN.  Many of the galaxies we observed with suppressed (PAH
8~$\mu$m)/160~$\mu$m ratios in their centres are objects that are
classified as containing AGN, although NGC~3184 is a notable
exception.  Another possibility is that the (PAH 8~$\mu$m)/160~$\mu$m
ratio drops in regions where a significant fraction of the
interstellar radiation field originates from evolved bulge stars.  As
explained above, PAHs may not be heated as strongly by single photons
in such radiation fields, and therefore 7.7~$\mu$m emission will be
inhibited.

\section{Conclusion}
\label{s_conclusions}

We have shown that the relation between PAH 8 and 24~$\mu$m emission
in this sample of spiral galaxies exhibit significant scatter on
scales of $\sim2$~kpc.  In particular, we have shown that the PAH
8~$\mu$m emission is relatively weak compared to 24~$\mu$m emission in
star-forming regions, but the PAH 8~$\mu$m emission is relatively
strong in the diffuse ISM.  In some cases, the (PAH
8~$\mu$m)/24~$\mu$m ratio may exhibit variations greater than a factor
of 2 at a given 24~$\mu$m surface brightness.  We argue that two
mechanisms are responsible for these variations: enhancements in the
24~$\mu$m emission relative to PAH emission in regions with very
strong radiation fields with high-energy photons and the corresponding
reduction of PAH 8~$\mu$m emission.

We have also shown that the PAH 8~$\mu$m emission is associated with
160~$\mu$m emission on scales of $\sim2$~kpc within these spiral
galaxies and that the (PAH 8~$\mu$m)/160~$\mu$m ratio appears to be a
function of 160~$\mu$m surface brightness.  The scatter in the
relation is only at the 10\%-20\% level for a given surface
brightness, and the intrinsic scatter measured in the
relation between the (PAH 8~$\mu$m)/160~$\mu$m ratio and 160~$\mu$m
surface brightness is notably lower than for the relation between the
(PAH 8~$\mu$m)/24~$\mu$m ratio and 24~$\mu$m surface brightness.
While the (PAH 8~$\mu$m)/160~$\mu$m ratio appears to decrease
monotonically with radius in some galaxies, the presence of peaks in
the (PAH 8~$\mu$m)/160~$\mu$m ratio corresponding to large scale
structure indicates that surface brightness may be a more important
factor, and statistical tests of the best fit lines between the (PAH
8~$\mu$m)/160~$\mu$m ratio and either 160~$\mu$m surface brightness or
radius showed that the dependence with 160~$\mu$m surface brightness
was more significant.  The strong correlation between the (PAH
8~$\mu$m)/160~$\mu$m ratio and 160~$\mu$m emission and the results
from the comparison of PAH 8 and 24~$\mu$m emission suggest that most
of the PAHs are located in the diffuse interstellar medium with the
dust grains that produce the majority of the 160~$\mu$m emission.  We
therefore suggest that the (PAH 8~$\mu$m)/160~$\mu$m ratio may be
indicative of the intensity of the interstellar radiation field that
heats the diffuse interstellar dust in these galaxies.

The results here indicate that PAH emission should be used very
cautiously as a tracer of star formation on kiloparsec scales if it
should be used at all.  Instead, the PAH emission may be more
indicative of the distribution of diffuse dust within nearby galaxies,
although PAH emission still may be affected by metallicity
\citep{eetal05, detal05, cetal07, eetal08}, and the results here
suggest that PAH emission might be inhibited in the diffuse ISM if the
radiation field is very high.  Follow-up observations with the
Herschel Space Observatory and the James Clerk Maxwell Telescope will
allow for studying the correlation between PAH and $\sim20$~K dust
emission on smaller spatial scales in nearby galaxies, thus placing
further constraints on the relation between PAHs and cool dust.

\section*{Acknowledgements}

This work was funded by STFC.  BTD was supported in part by NSF grant
AST04-06883.  AL is supported in part by the {\it Spitzer} Theory
Programs and NSF grant AST 07-07866.

\label{lastpage}

\end{document}